\documentstyle{article}
\input epsf
\begin{document}


\title{Critical phenomena in gravitational collapse}

\author{Carsten Gundlach \\
Max-Planck-Institut f\"ur Gravitationsphysik
(Albert-Einstein-Institut) \\ 
Schlaatzweg 1, 14473 Potsdam, Germany}

\date{9 Jan 1998, revised 26 Jan 1998}

\maketitle


\begin{abstract}

As first discovered by Choptuik, the black hole threshold in the space
of initial data for general relativity shows both surprising structure
and surprising simplicity. Universality, power-law scaling of the
black hole mass, and scale echoing have given rise to the term
``critical phenomena''. They are explained by the existence of exact
solutions which are attractors within the black hole threshold, that
is, attractors of codimension one in phase space, and which are
typically self-similar. This review gives an introduction to the
phenomena, tries to summarize the essential features of what is
happening, and then presents extensions and applications of this basic
scenario. Critical phenomena are of interest particularly for creating
surprising structure from simple equations, and for the light they
throw on cosmic censorship. They may have applications in quantum
black holes and astrophysics.

\end{abstract}


\tableofcontents


\section{Introduction}


In 1987 Christodoulou, who was then (and still is) studying the
spherically symmetric Einstein-scalar model analytically
\cite{Christodoulou1,Christodoulou2,Christodoulou3,Christodoulou4,Christodoulou5}
suggested to Matt Choptuik, who was investigating the same system
numerically, the following question \cite{Choptuik94}: Consider a
generic smooth one-parameter family of initial data, such that for
large values of the parameter $p$ a black hole is formed, and no black
hole is formed for small $p$. If one makes a bisection search for the
critical value $p_*$ where a black hole is just formed, does it have
finite or infinitesimal mass?  After developing advanced numerical
methods for this purpose, Choptuik managed to give highly convincing
numerical evidence that the mass is infinitesimal. Moreover he found
two totally unexpected phenomena: The first is the now famous scaling
relation
\begin{equation}
\label{power_law}
M \simeq C \, (p-p_*)^\gamma
\end{equation}
for the black hole mass $M$ in the limit $p\simeq p_*$ (but $p>p_*$),
where the constant $\gamma$ is the same for all such one-parameter
families. (Choptuik found $\gamma\simeq 0.37$.) The second is the
appearance of a highly complicated, scale-periodic solution for $p
\simeq p_*$, which is again the same for all initial data as long as
they are near the limit of black hole formation. The logarithmic scale
period of this solution, $\Delta\simeq 3.44$, is a second
dimensionless number coming out of the blue.

Until then most relativists would have assigned numerical work the
role of providing quantitative details of phenomena that were already
understood qualitatively, noticeably in astrophysical
applications. Here, numerical relativity
provided an important qualitative input into mathematical relativity
and gave rise to a new research field. Similar phenomena to
Choptuik's results were quickly found in other systems too, suggesting
that they were limited neither to scalar field matter nor to spherical
symmetry. Many researchers were intrigued by the appearance of a
complicated ``echoing'' structure, and the two mysterious
dimensionless numbers, in the critical solution. Later it was realized
that critical phenomena also provide a natural route to naked
singularities, and this has linked critical phenomena to the
mainstream of research in mathematical relativity.  Purely analytical
approaches, however, have not been successful so far, and most of what
is understood in critical phenomena is based on a mixture of
analytical and numerical work.  Scale-invariance, universality and
power-law behavior suggest the name critical phenomena. A connection
with the renormalisation group in partial differential equations has
been established in hindsight, but has not yet provided fresh
input. The connection with the renormalisation group in statistical
mechanics is even more tenuous, limited to approximate scale
invariance, but not extending to the presence of a statistical
ensemble.

In our presentation we combine a phenomenological with a
systematic approach.  In order to give the reader not familiar with
Choptuik's work a flavor of how complicated phenomena arise from
innocent-looking PDEs, we describe his results in some detail,
followed by a review of the work of Coleman and Evans on
critical phenomena in perfect fluid collapse, which appeared a year
later. (The important paper of Abrahams and Evans, historically the
first paper after Choptuik's, is reviewed in the context of
non-spherically symmetric systems.)

After this phenomenological opening, we systematically explain the key
features echoing, universality and scaling in a coherent scenario
which has emerged over time, with key terminology borrowed from
dynamical systems and renormalisation group theory. This picture is
partly qualitative, but has been underpinned by successful
semi-analytic calculations of Choptuik's (and other) critical
solutions and the critical exponent $\gamma$ to high
precision. Semi-analytic here means that although an analytic solution
is impossible, the numerical work solves a simplified problem, for
example reducing a PDE to an ODE. In this context we introduce the
relativistic notions of scale-invariance and scale-periodicity, define
the concept of a critical solution, and sketch the calculation of the
critical exponent.

In the following section we present extensions of this basic
scenario. This presentation is again systematic, but to also give the
phenomenological point of view, the section starts with a tabular
overview of the matter models in which critical phenomena have been
studied so far. Extensions of the basic scenario include more realistic
matter models, critical phenomena with a mass gap, the study of the
global structure of the critical spacetime itself, and black holes
with charge and mass.

In a final section that could be titled ``loose ends'', we group
together approaches to the problem that have failed or are as yet at a
more speculative stage. This section also reviews some detailed work
on the quantum aspects of critical collapse, based on various toy
models of semiclassical gravity.

Previous short review papers include Horne \cite{Horne_MOG}, Bizo\'n
\cite{Bizon} and Gundlach \cite{Gundlach_Banach}. Choptuik is 
preparing a longer review paper \cite{Choptuik_review}. For an
interesting general review of the physics of scale-invariance, see
\cite{Wilson}. 


\section{A look at the phenomena}


\subsection{The spherically symmetric scalar field} \label{Choptuik_results}


The system in which Christodoulou and Choptuik studied gravitational
collapse in detail was the spherically symmetric massless, minimally
coupled scalar field. It has the advantage of simplicity, and the
scalar radiation propagating at the speed of light mimics
gravitational waves within spherical symmetry. The Einstein equations
are
\begin{equation}
\label{scalar_stress_energy}
G_{ab} = 8 \pi \left(\nabla_a \phi \nabla_b \phi - {1\over 2} g_{ab}
\nabla_c \phi \nabla^c \phi\right)
\end{equation}
and the matter equation is
\begin{equation}
\nabla_a \nabla^a \phi = 0.
\end{equation}
Note that the matter equation of motion is contained within the
contracted Bianchi identities. Choptuik chose Schwarzschild-like
coordinates 
\begin{equation}
\label{tr_metric}
ds^2 = - \alpha^2(r,t) \, dt^2 + a^2(r,t) \, dr^2 + r^2 \, d\Omega^2,
\end{equation}
where $d\Omega^2 = d\theta^2 + \sin^2\theta \, d\varphi^2$ is the
metric on the unit 2-sphere. This choice of coordinates is defined by
the radius $r$ giving the surface of 2-spheres as $4\pi r^2$, and by
$t$ being orthogonal to $r$. One more condition is required to fix the
coordinate completely. Choptuik chose $\alpha=1$ at $r=0$, so that $t$
is the proper time of the central observer.

In the auxiliary variables
\begin{equation}
\Phi = \phi_{,r}, \qquad \Pi={a\over\alpha} \phi_{,t},
\end{equation}
the wave equation becomes a first-order system,
\begin{eqnarray}
\label{wave}
\Phi_{,t} & = &  \left({\alpha\over a}\Pi\right)_{,r}, \\
\Pi_{,t} & = & {1\over r^2} \left({\alpha\over a} \Phi\right)_{,r}.
\end{eqnarray}
In spherical symmetry there are four algebraically independent
components of the Einstein equations. Of these, one is proportional to
derivatives of the other and can be disregarded. The other three
contain only first derivatives of the metric, namely $a_{,t}$,
$a_{,r}$ and $\alpha_{,r}$. Choptuik chose to use the equations giving
$a_{,r}$ and $\alpha_{,r}$ for his numerical scheme, so that only the
scalar field is evolved, but the two metric coefficients are
calculated from the matter at each new time step. (The main advantage
of such a numerical scheme is its stability.) These two equations are
\begin{eqnarray} 
\label{da_dr}
{1\over a} a_{,r} + {a^2 -1 \over 2r} - 2\pi r (\Pi^2 + \Phi^2) & = &
0, \\
\label{dalpha_dr}
{1\over \alpha} \alpha_{,r} - {1\over a} a_{,r} - {a^2 -1 \over 2r} &
= & 0,
\end{eqnarray}
and they are, respectively, the Hamiltonian constraint and the slicing
condition. These four first-order equations totally describe the
system. For completeness, we also give the remaining Einstein
equation,
\begin{equation}
\label{da_dt}
{1\over \alpha} a_{,t} =  2\pi r (\Pi^2 - \Phi^2).
\end{equation}

The free data for the system are the two functions $\Pi(r)$ and
$\Phi(r)$. (In spherical symmetry, there are no physical degrees of
freedom in the gravitational field.) Choptuik investigated many
one-parameter families of such data by evolving the data for many
values each of the parameter, say $p$. Simple examples of such
families are $\Phi(r)=0$ and a Gaussian for $\Pi(r)$, with the
parameter $p$ taken to be either the amplitude of the Gaussian, with
the width and center fixed, or the width, with position and amplitude
fixed, or the position, with width and amplitude fixed. It is
plausible that for the amplitude sufficiently small, with width and
center fixed, the scalar field will disperse, and for sufficiently
large amplitude will form a black hole, with similar behavior for many
generic parameters.  This is difficult to prove in
generality. Christodoulou showed for the spherically symmetric scalar
field system that data sufficiently weak in a well-defined way evolve
to a Minkowski-like spacetime \cite{Christodoulou3}, and that a class
of sufficiently strong data forms a black hole \cite{Christodoulou2}.

But what happens in between?  Choptuik found that in all families of
initial data he could make arbitrarily small black holes by
fine-tuning the parameter $p$ close to the black hole threshold. An
important fact is that there is nothing visibly special to the black
hole threshold. One cannot tell that one given data set will form a
black hole and another one infinitesimally close will not, short of
evolving both for a sufficiently long time. Fine-tuning is then a
heuristic procedure, and effectively proceeds by bisection: Starting
with two data sets one of which forms a black hole, try a third one
in between along some one-parameter family linking the two, drop one
of the old sets and repeat.

With $p$ closer to $p_*$, the spacetime varies on ever smaller
scales. The only limit was numerical resolution, and in order to push
that limitation further away, Choptuik developed special numerical
techniques that recursively refine the numerical grid in spacetime
regions where details arise on scales too small to be resolved
properly. In the end, Choptuik could determine $p_*$ up to a relative
precision of $10^{-15}$, and make black holes as small as $10^{-6}$
times the ADM mass of the spacetime. The power-law scaling
(\ref{power_law}) was obeyed from those smallest masses up to black
hole masses of, for some families, $0.9$ of the ADM mass, that is,
over six orders of magnitude
\cite{Choptuik94}. There were no families of initial data which did
not show the universal critical solution and critical
exponent. Choptuik therefore conjectured that $\gamma$ is the same for
all one-parameter families, and that the approximate scaling law holds
ever better for arbitrarily small $p-p_*$.

I would suggest reformulating this conjecture in a different
manner. Let us first consider a finite-dimensional subspace of the
space of initial data, with coordinates $p_i$ on it. The subspace of
Gaussian data for both $\Phi$ and $\Pi$ for example is
6-dimensional. We could choose the amplitudes, centers and widths of
the two Gaussians as coordinates, but any six smooth functions of
these could also serve as coordinates. Various one-parameter families
only serve as probes of this one 6-dimensional space. They
indicate that there is a smooth hypersurface in this space which
divides black hole from non-black hole data. Let $P(p_i)$ be any
smooth coordinate function on the space so that $P(p_i)=0$ is the
black hole threshold. Then, for any choice of $P(p_i)$, there is a
second smooth function $C(p_i)$ on the space so that the black hole
mass as a function of the space is given as
\begin{equation}
M = \theta(P) \, C  P^\gamma.
\end{equation}
In words, the entire unsmoothness at the black hole threshold is
captured by the one critical exponent.  One can now formally go over
from a finite-dimensional subspace to the infinite-dimensional space
of initial data. Note that $\Phi$ and $\Pi$ must be square-integrable
for the spacetime to be asymptotically flat, and therefore the initial
data space has a countable basis. In this view, it is difficult to see
how different one-parameter families could have different values of
$\gamma$. It also shows that the critical exponent is not an effect of
a bad parameterization.

Clearly a collapse spacetime which has ADM mass 1, but settles down to
a black hole of mass (for example) $10^{-6}$ has to show structure on
very different scales. The same is true for a spacetime which is as
close to the black hole threshold, but on the other side: the scalar
wave contracts until curvature values of order $10^{12}$ are reached
in a spacetime region of size $10^{-6}$ before it starts to
disperse. Choptuik found that all near-critical spacetimes, for all
families of initial data, look the same in an intermediate region,
that is they approximate one universal spacetime, which is also called
the critical solution. This spacetime is scale-periodic in the sense
that there is a value $t_*$ of $t$ such that when we shift the origin
of $t$ to $t_*$, we have
\begin{equation}
\label{tr_scaling}
Z(r,t) = Z\left(e^{n\Delta}r,e^{n\Delta}t\right)
\end{equation}
for all integer $n$ and for $\Delta\simeq 3.44$, and where $Z$ stands
for any one of $a$, $\alpha$ or $\phi$ (and therefore also for $r\Pi$
or $r\Phi$). The accumulation point $t_*$ depends on the family, but
the scale-periodic part of the near-critical solutions does not.

This result is sufficiently surprising to formulate it once more in a
slightly different manner. Let us replace $r$ and $t$ by a pair of
auxiliary variables such that one of them retains a dimension, while
the other is dimensionless. A simple example is (after shifting the
origin of $t$ to $t_*$)
\begin{equation}
\label{x_tau}
x = -{r \over t}, \quad \tau = - \ln\left(-{t\over L}\right),
\quad t<0.
\end{equation}
(As a matter of convention, $t$ has been assumed negative so that it
increases towards the accumulation point at $t=r=0$. Similarly, $\tau$
has been defined so that it increases with increasing $t$.)
Choptuik's observation, expressed in these coordinates, is that in any
near-critical solution there is a space-time region where the fields
$a$, $\alpha$ and $\phi$ are well approximated by their values in a
universal solution, as
\begin{equation}
Z(x,\tau) \simeq Z_*(x,\tau),
\end{equation}
where the fields $a_*$, $\alpha_*$ and $\phi_*$ of the critical solution 
have the property
\begin{equation}
Z_*(x,\tau+\Delta) = Z_*(x,\tau).
\end{equation}
The dimensionful constants $t_*$ and $L$ depend on the one-parameter
family of solutions, but the dimensionless critical fields $a_*$,
$\alpha_*$ and $\phi_*$, and in particular their dimensionless period
$\Delta$, are universal. A slightly supercritical and a slightly
subcritical solution from the same family (so that $L$ and $t_*$ are
the same) are practically indistinguishable until they have reached a
very small scale where the one forms an apparent horizon, while the
other starts dispersing. Not surprisingly, this scale is the same as
that of the black hole (if one is formed), and so we have for the
range $\Delta\tau$ of $\tau$ on which a near-critical solution
approximates the universal one
\begin{equation}
\label{duration}
\Delta\tau \simeq \gamma \ln|p-p_*| + const
\end{equation}
and for the number $N$ of scaling ``echos'' that are seen,
\begin{equation}
N \simeq \Delta^{-1} \gamma \ln|p-p_*| + {\rm const.}
\end{equation}
Note that this holds for both supercritical and subcritical solutions.

Choptuik's results have been repeated by a number of other authors.
Gundlach, Price and Pullin \cite{GPP2} could verify the mass scaling
law with a relatively simple code, due to the fact that it holds even
quite far from criticality. Garfinkle \cite{Garfinkle} used the fact
that recursive grid refinement in near-critical solutions is not
required in arbitrary places, but that all refined grids are centered
on $(r=0,t=t_*)$, in order to simulate a simple kind of mesh
refinement on a single grid in double null coordinates: $u$ grid lines
accumulate at $u=0$, and $v$ lines at $v=0$, with $(v=0,u=0)$ chosen
to coincide with $(r=0,t=t_*)$. Hamad\'e and Stewart
\cite{HamadeStewart} have written a complete mesh refinement algorithm
based on a double null grid (but coordinates $u$ and $r$), and report
even higher resolution than Choptuik. Their coordinate choice also
allowed them to follow the evolution beyond the formation of an
apparent horizon.


\subsection{The spherically symmetric perfect fluid}


The scale-periodicity, or echoing, of the scalar field critical
solution was a new phenomenon in general relativity, and initial
efforts at understanding concentrated there. Evans however realized
that scale-echoing was only a more complicated form of
scale-invariance, that the latter was the key to the problem, and
moreover that it could be expected to arise in a different matter
model, namely a perfect fluid. Evans knew that scale-invariant, or
self-similar, solutions arise in fluid dynamics problems (without
gravity) when there are two very different scales in the initial
problem (for example an explosion with high initial density into a
thin surrounding fluid \cite{BarenblattZeldovich}), and that such
solutions play the role of an intermediate asymptotic in the
intermediate density regime \cite{Barenblatt}.

Evans and Coleman \cite{EvansColeman} therefore made a perfect fluid
ansatz for the matter,
\begin{equation}
\label{fluid_stress_energy}
G_{ab} = 8 \pi \left[(p+\rho) u_a u_b + p g_ab\right]
\end{equation}
where $u^a$ is the 4-velocity, $\rho$ the density and $p$ the
pressure. As for the scalar field, the matter equations of motion are
equivalent to the conservation of matter energy-momentum. The
only equation of state without an intrinsic scale is $p=k\rho$, with
$k$ a constant. This was desirable in order to allow for a
scale-invariant solution like that of Choptuik. Evans and Coleman
chose $k=1/3$ because it is the equation of state of radiation (or
ultra-relativistic hot matter). They made the same coordinate choice in
spherical symmetry as Choptuik, and evolved one-parameter families of
initial data. They found a universal intermediate attractor, and
power-law scaling of the black hole mass, with a universal critical
exponent of $\gamma\simeq 0.36$. (To anticipate, the coincidence of
the value with that for the scalar field is now believed to be
accidental.) The main difference was that the universal solution is
not scale-periodic but scale-invariant: it has the continuous symmetry
(after shifting the origin of $t$ to $t_*$)
\begin{equation} 
Z_*(r,t) = Z_*\left(-{r\over t}\right) = Z(x)
\end{equation}
where $Z$ now stands for the metric coefficients $a$ and $\alpha$ (as
in the scalar field case) and the dimensionless matter variables
$t^2\rho$ and $u^r$. We shall
discuss this symmetry in more detail below.

Independently, Evans and Coleman made a scale-invariant ansatz for the
critical solution, which transforms the PDE problem in $t$ and $r$
into an ODE problem in the one independent variable $x$. They then
posed a nonlinear boundary value problem by demanding regularity
at the center $x=0$ and at the past sound cone $x=x_0$ of the point
$(t=r=0)$, where a generic self-similar solution would be singular.
The sound cone referred to here and below is a characteristic of the
matter equations of motion. It is made up of the characteristic curves
which are also homothetic curves ($x={\rm const.}$).  The past {\it
light} cone of $(t=r=0)$ plays no role in the spherically symmetric
perfect fluid critical solution because in spherical symmetry there
are no propagating gravitational degrees of freedom. It does play a
role in the spherically symmetric scalar field critical solutions
however, because there it is also characteristic of the scalar field
matter.

The regularity condition at the center $x=0$ is local flatness, or
$a=1$ in the coordinates (\ref{tr_metric}). The regularity condition
at $x=x_0$ is the absence of a shock wave. I believe that both
conditions are equivalent to demanding analyticity, and that $x=0$ and
$x=x_0$ are ``regular singular points'' of the ODE system, although
this remains to be shown by a suitable change of variables. 
The solutions of this boundary value problem form a discrete family.

The simplest solution of the boundary value problem coincides
perfectly with the intermediate asymptotic that is found in the
collapse simulations, arising from generic data. It is really this
coincidence that justifies the boundary value problem posed by Evans
and Coleman. At an intuitive level, however, one could argue that the
critical solution should be smooth because it arises as an
intermediate asymptotic from smooth initial data. In a contrasting
opinion, Carr and Henriksen \cite{CarrHenriksen} claim that the
perfect fluid critical solution should obey a certain global condition
(the ``particle horizon'' and the ``event horizon'' of the spacetime
coincide) that can be interpreted as the solution being a marginal
black hole.  In order to impose this condition, they need one more
free parameter in the space of CSS solutions, and obtain it by not
imposing analyticity at the past sound cone, where their candidate
critical solution has a shock.

Evans and Coleman also suggested that an analysis of the linear
perturbations of the critical solution would give an ``estimate'' of
the critical exponents. This program was carried out for the $k=1/3$
perfect fluid by Koike, Hara and Adachi
\cite{KoikeHaraAdachi} and for other values of $k$ by Maison
\cite{Maison}, to high precision. 


\section{The basic scenario}


\subsection{Scale-invariance, self-similarity, and homothety}


The critical solution found by Choptuik
\cite{Choptuik91,Choptuik92,Choptuik94} for the spherically symmetric
scalar field 
is scale-periodic, or discretely self-similar (DSS), and the critical
solution found by Evans and Coleman
\cite{EvansColeman} is scale-invariant, or continuously
self-similar (CSS). We begin with the continuous symmetry because it
is simpler. In Newtonian physics, a solution $Z$ is self-similar if it
is of the form
\begin{equation}
Z(\vec x, t) = Z\left[{\vec x\over f(t)}\right]
\end{equation}
If the function $f(t)$ is derived from dimensional considerations
alone, one speaks of self-similarity of the first kind. An example is
$f(t)=\sqrt{\lambda t}$ for the diffusion equation $Z_{,t}=\lambda
Z_{,xx}$. In more complicated equations, the limit of self-similar
solutions can be singular, and $f(t)$ may contain additional
dimensionful constants (which do not appear in the field equation) in
terms such as $(t/L)^\alpha$, where $\alpha$, called an anomalous
dimension, is not determined by dimensional considerations but through
the solution of an eigenvalue problem \cite{Barenblatt}.  For now, we
concentrate on self-similarity of the first kind.

A continuous self-similarity of the spacetime in GR corresponds to the
existence of a homothetic vector field $\xi$, defined by the property
\cite{CahillTaub} 
\begin{equation}
\label{homothetic_metric}
{\cal L}_\xi g_{ab} = 2 g_{ab}.
\end{equation}
(This is a special type of conformal Killing vector, namely one with
constant coefficient on the right-hand side. The value of
this constant coefficient is conventional, and can be set equal to 2 by
a constant rescaling of $\xi$.) From (\ref{homothetic_metric}) it
follows that
\begin{equation}
\label{homothetic_curvature}
{\cal L}_\xi {R^a}_{bcd} = 0,
\end{equation}
and therefore
\begin{equation}
\label{homothetic_matter}
{\cal L}_\xi G_{ab} = 0,
\end{equation}
but the inverse does not hold: the Riemann tensor and the metric need
not satisfy (\ref{homothetic_curvature}) and (\ref{homothetic_metric})
if the Einstein tensor
obeys (\ref{homothetic_matter}). If the matter is a perfect fluid
(\ref{fluid_stress_energy}) it follows from (\ref{homothetic_metric}),
(\ref{homothetic_matter}) and the Einstein equations that
\begin{equation}
{\cal L}_\xi u^a = -u^a, \quad {\cal L}_\xi \rho = -2\rho, \quad
{\cal L}_\xi p = -2p.
\end{equation}
Similarly, if the matter is a free scalar field $\phi$
(\ref{scalar_stress_energy}), it follows that
\begin{equation}
\label{scalar_CSS}
{\cal L}_\xi \phi=\kappa, 
\end{equation}
where $\kappa$ is a constant.

In coordinates $x^\mu=(\tau,x^i)$ adapted to the homothety, the metric
coefficients are of the form
\begin{equation}
\label{CSS_coordinates}
g_{\mu\nu}(\tau,x^i) = e^{-2\tau} \tilde g_{\mu\nu}(x^i)
\end{equation}
where the coordinate $\tau$ is the negative logarithm of a
spacetime scale, and the remaining three coordinates $x^i$ are
dimensionless. In these coordinates, the homothetic vector field is
\begin{equation}
\label{xi_in_coordinates}
\xi = - {\partial\over\partial\tau}.
\end{equation}
The minus sign in both equations (\ref{CSS_coordinates}) and
(\ref{xi_in_coordinates}) is a convention we have chosen so that
$\tau$ increases towards smaller spacetime scales. For the critical
solutions of gravitational collapse, we shall later choose surfaces of
constant $\tau$ to be spacelike (although this is not possible
globally), so that $\tau$ is the time coordinate as well as the scale
coordinate. Then it is natural that $\tau$ increases towards the
future, that is towards smaller scales.

As an illustration, the CSS scalar field in these coordinates would be
\begin{equation}
\phi=f(x)+\kappa\tau,
\end{equation}
with $\kappa$ a constant.

The generalization to a discrete self-similarity is
obvious in these coordinates, and was made in \cite{Gundlach_Chop2}:
\begin{equation}
\label{DSS_coordinates}
g_{\mu\nu}(\tau,x^i) = e^{-2\tau} \tilde g_{\mu\nu}(\tau,x^i), \quad
\hbox{where} \quad \tilde g_{\mu\nu}(\tau,x^i) = \tilde 
g_{\mu\nu}(\tau+\Delta,x^i).
\end{equation}
The conformal metric $\tilde g_{\mu\nu}$ does now depend on $\tau$,
but only in a periodic manner. Like the continuous symmetry, the
discrete version has a geometric formulation
\cite{Gibbonspc}: A spacetime is discretely self-similar if 
there exists a discrete diffeomorphism $\Phi$ and a real constant
$\Delta$ such that
\begin{equation} 
\label{DSS_geometric}
\Phi^* g_{ab} = e^{2\Delta} g_{ab},
\end{equation}
where $\Phi^* g_{ab}$ is the pull-back of $g_{ab}$ under the
diffeomorphism $\Phi$. This is our definition of discrete
self-similarity (DSS). It can be obtained formally from
(\ref{homothetic_metric}) by integration along $\xi$ over an interval
$\Delta$ of the affine parameter. Nevertheless, the definition is
independent of any particular vector field $\xi$. One simple
coordinate transformation that brings the Schwarzschild-like
coordinates (\ref{tr_metric}) into this form, with the periodicity in
$\tau$ equivalent to the scaling property (\ref{tr_scaling}), was
given above in Eqn. (\ref{x_tau}), as one easily verifies by
substitution.  The most general ansatz for the massless scalar field
compatible with DSS is
\begin{equation}
\label{scalar_DSS}
\phi=f(\tau,x^i) + \kappa \tau, \quad {\rm where} \quad
f(\tau,x^i)=f(\tau+\Delta,x^i)
\end{equation}
with $\kappa$ a constant.

It should be stressed here that the coordinate systems adapted to CSS
(\ref{CSS_coordinates}) or DSS (\ref{DSS_coordinates}) form large
classes, even in spherical symmetry. One can fix the surface $\tau=0$
freely, and can introduce any coordinates $x^i$ on it. In particular,
in spherical symmetry, $\tau$-surfaces can be chosen to be spacelike,
as for example defined by (\ref{tr_metric}) and (\ref{x_tau}) above,
and in this case the coordinate system cannot be global (in the
example, $t<0$).  Alternatively, one can find global coordinate
systems, where $\tau$-surfaces must become spacelike at large $r$, as
in the coordinates (\ref{global_CSS}). Moreover, any such coordinate
coordinate system can be continuously deformed into one of the same
class.

As an aside, we mention that self-similarity of the second kind in
general relativity was studied by Carter and Henriksen
\cite{CarterHenriksen} and Coley \cite{Coley}. 
The connection with the Newtonian definition is that
space and time are rescaled in different ways. To make this a
covariant notion one needs a preferred timelike congruence. The
4-velocity $u^a$ of a perfect fluid is a natural candidate. The metric
$g_{ab}$ can then be decomposed into space and time as $g_{ab}=-u_a
u_b +h_{ab}$. The homothetic scaling (\ref{homothetic_metric}) is
replaced by
\begin{equation}
{\cal L}_\xi h_{ab} = 2 h_{ab}, \quad {\cal L}_\xi u_a = C u_a, 
\end{equation}
with $C\ne 1$. This kind of self-similarity has not to date
been found in critical collapse.  In a possible source of confusion, Evans
and Coleman \cite{EvansColeman} use the term ``self-similarity of the
second kind'', because they define their self-similar coordinate $x$
as $x=r/f(t)$, with $f(t)=t^n$. Nevertheless, the spacetime they
calculate is homothetic, that is, self-similar of the first kind
according to the terminology of Carter and Henriksen. The difference
is only a coordinate transformation: the $t$ of
\cite{EvansColeman} is not proper time at the origin, but what would
be proper time at infinity if the spacetime was truncated at finite
radius and matched to an asymptotically flat exterior
\cite{Evanspc}.

There is a large body of research on spherically symmetric
self-similar solutions. A detailed review is
\cite{CarrColey}. Here we should mention only that
perfect fluid spherically symmetric self-similar solutions have been
examined by Bogoyavlenskii \cite{Bogoyavlenskii}, Foglizzo and
Henriksen
\cite{FoglizzoHenriksen}, Bicknell and Henriksen
\cite{BicknellHenriksen} and Ori and Piran \cite{OriPiran}. Scalar
field spherically symmetric CSS solutions were examined by Brady
\cite{Brady_CSS_scalar}. In these papers, the Einstein equations are
reduced to an ODE system by the self-similar spherically symmetric
ansatz, which is then discussed as a dynamical system. It is often
difficult to regain the spacetime picture from the phase space
picture. In particular, it is not clear which solution in these
classifications is the critical solution found in perfect fluid
collapse simulations, and constructed through a CSS ansatz, by Evans
and Coleman \cite{EvansColeman} (but see
\cite{CarrHenriksen}). It is also unclear why the scalar
field DSS critical solution has $\kappa=0$ in Eqn. (\ref{scalar_DSS}).


\subsection{Gravity regularizes self-similar matter} \label{section_regular}


It is instructive to consider the self-similar solutions of a simple
matter field, the massless scalar field, in spherical symmetry without
gravity. The general solution of the spherically symmetric wave
equation is of course
\begin{equation} 
\phi(r,t) = r^{-1}\left[f(t+r)-g(t-r)\right],
\end{equation}
where $f(z)$ and $g(z)$ are two free functions of one variable ranging
from $-\infty$ to $\infty$. $f$ describes ingoing and $g$ outgoing
waves. Regularity at the center $r=0$ for all $t$ requires $f(z)=g(z)$
for $f(z)$ a smooth function. Physically this means that ingoing waves
move through the center and become outgoing waves. Now we transform to
new coordinates $x$ and $\tau$ defined by
\begin{equation}
\label{global_CSS}
r = e^{-\tau} \cos x, \qquad t = e^{-\tau} \sin x,
\end{equation}
and with range $-\infty<\tau<\infty$, $-\pi/2\le x\le \pi/2$. These
coordinates are adapted to self-similarity, but unlike the $x$ and
$\tau$ introduced in (\ref{x_tau}) they cover all of Minkowski space
with the exception of the point $(t=r=0)$. The general solution of the
wave equation for $t>r$ can formally be written as
\begin{eqnarray}
\phi(r,t)=\phi(x,\tau) & = & (\tan x + 1)F_+\left[\ln(\sin x + \cos x) 
- \tau\right] \nonumber \\
& - & (\tan x - 1)G_+\left[\ln(\sin x - \cos x) - \tau\right],
\end{eqnarray}
through the substitution $f(z)/z=F_+(\ln z)$ and $g(z)/z=G_+(\ln z)$
for $z>0$. Similarly, we
define $f(z)/z=F_-[\ln (-z)]$ and $g(z)/z=G_-[\ln (-z)]$ for $z<0$ to
cover the sectors $|t|<r$ and $t<-r$. Note that $F_+(z)$ and $F_-(z)$
together contain the same information as $f(z)$.

The condition for regularity at $r=0$ for all $t>0$ is once more
$F_+(z)=G_+(z)$, but we can now also read off that the condition for
continuous self-similarity $\phi=\phi(x)$ translates into $F_+={\rm
const.}$, $G_+={\rm const.}$. Discrete self-similarity with scale
periodicity $\Delta$, or $\phi(x,\tau)=\phi(x,\tau+\Delta)$ translates
into $F_+(z)=F_+(z+\Delta)$ and $G_+(z)=G_+(z+\Delta)$. Any
self-similar solution is singular at $t=r$ unless $G_+=0$. Similar
conclusions are obtained for the sectors $|t|<r$ and $t<-r$.
We conclude that a self-similar solution (continuous or discrete) is
either zero everywhere, or else it is regular in at most {\it one} of
three places: at the center $r=0$ for $t\ne 0$, at the past light cone
$t=-r$, or at the future light cone $t=r$. (These three cases
correspond to $F_+=G_+$ and $F_-=G_-$, $F_+=F_-=0$, and $G_+=G_-=0$,
respectively.) We conjecture that other simple matter fields, such as
the perfect fluid, show similar behavior.

The presence of gravity changes this singularity structure
qualitatively. Dimensional analysis applied to the metric
(\ref{CSS_coordinates}) or (\ref{DSS_coordinates}) shows that
$\tau=\infty$ [the point $(t=r=0)$] is now a curvature singularity
(unless the self-similar spacetime is Minkowski). But elsewhere, the
solution can be more regular. There is a one-parameter family of exact
spherically symmetric scalar field solutions found by Roberts
\cite{Roberts} that
is regular at both the future and past light cone of the singularity,
not only at one of them. (It is singular at the past and future branch
of $r=0$.) The only solution without gravity with this property
is $\phi=0$. The Roberts solution will be discussed in more detail in
section \ref{section_singularity} below.

Similarly, the scale-invariant or scale-periodic solutions found in
near-critical collapse simulations are regular at both the past branch
of $r=0$ and the past light cone (or sound cone, in the case of the
perfect fluid). Once more, in the absence of
gravity only the trivial solution has this property. 

I have already argued that the critical solution must be as smooth on
the past light cone as elsewhere, as it arises from the collapse of
generic smooth initial data. No lowering of differentiability or other
unusual behavior should take place before a curvature singularity
arises at the center. As Evans first realized, this requirement turns
the scale-invariant or scale-periodic ansatz into a boundary value
problem between the past branch of $r=0$ and the past sound cone, that
is, roughly speaking, between $x=0$ and $x=1$.

In the CSS ansatz in spherical symmetry suitable for the perfect
fluid, all fields depend only on $x$, and one obtains an ODE boundary
value problem. In a scale-periodic ansatz in spherical symmetry, such
as for the scalar field, all fields are periodic in $\tau$, and one
obtains a 1+1 dimensional hyperbolic boundary value problem on a
coordinate square, with regularity conditions at, say, $x=0$ and
$x=1$, and periodic boundary conditions at $\tau=0$ and
$\tau=\Delta$. Well-behaved numerical solutions of these problems have
been obtained, with numerical evidence that they are locally unique,
and they agree well with the universal solution that emerges in
collapse simulations (references are given in the column ``Critical
solution'' of Table 1). It remains an open mathematical problem to
prove existence and (local) uniqueness of the solution defined by
regularity at the center and the past light cone.

One important technical detail should be mentioned here. In the curved
solutions, the past light cone of the singularity is not in general
$r=-t$, or $x=1$, but is given by $x=x_0$, or in the case of
scale-periodicity, by $x=x_0(\tau)$, with $x_0$ periodic in $\tau$
and initially unknown. The same problem arises for the sound cone. It
is convenient to make the coordinate transformation 
\begin{equation}
\bar x = {x \over x_0(\tau)}, \qquad \bar \tau = {2\pi\over
\Delta}\tau,
\end{equation}
so that the sound
cone or light cone is by definition at $\bar x =1$, while the origin
is at $\bar x =0$, and so that the period in $\bar\tau$ is now always
$2\pi$. In the DSS case the periodic function
$x_0(\bar\tau)$ and the constant $\Delta$ now appear explicitly in the
field equations, and they must be solved for as nonlinear eigenvalues.
In the CSS case, the constant $x_0$ appears, and must be solved for as
a nonlinear eigenvalue.

As an example for a DSS ansatz, we give the equations for the
spherically symmetric massless scalar field in the coordinates
(\ref{x_tau}) adapted to self-similarity and in a form ready for
posing the boundary value problem. (The equations of
\cite{Gundlach_Chop1} have been adapted to the notation of this
review.) We introduce the first-order matter variables
\begin{equation}
\label{Xpm}
X_\pm = \sqrt{2\pi} r\left({\phi_{,r}\over a}\pm {\phi_{,t}\over
\alpha}\right),
\end{equation}
which describe ingoing and outgoing waves. It is also useful to
replace $\alpha$ by
\begin{equation}
D = \left(1-{\Delta\over2\pi}{d\ln x_0\over
d\bar\tau}\right) {xa\over \alpha}
\end{equation}
as a dependent variable.  In the scalar field wave equation (\ref{wave})
we use the Einstein equations (\ref{dalpha_dr}) and (\ref{da_dt}) to
eliminate $a_{,t}$ and $\alpha_{,r}$, and obtain
\begin{equation}
\bar x {\partial X_\pm \over \partial \bar x}  =  (1 \mp D)^{-1}
\left\{
\left[{1\over2}(1-a^2)-a^2X_\mp^2\right] X_\pm - X_\mp 
\pm D 
\left({\Delta\over2\pi}-{d\ln x_0\over d\bar\tau}\right)^{-1}
{\partial X_\pm\over\partial \bar\tau}\right\}.
\end{equation}
The three Einstein equations (\ref{da_dr},\ref{dalpha_dr},\ref{da_dt})
become
\begin{eqnarray}
{\bar x\over a}{\partial a\over\partial\bar x} &  = & {1\over2}(1-a^2) 
+{1\over2} a^2 (X_+^2+X_-^2), \\
{\bar x\over D}{\partial D\over\partial\bar x} & = & 2 - a^2, \\ 
0 & = & (1-a^2) 
+a^2 (X_+^2+X_-^2) - a^2 D^{-1}(X_+^2-X_-^2) \nonumber \\
&&  + \left({\Delta\over2\pi}-{d\ln x_0\over d\bar\tau}\right)^{-1}
{2\over a} {\partial a\over \partial \bar\tau} .
\end{eqnarray}
As suggested by the format of the equations, they can be treated as
four evolution equations in $\bar x$ and one constraint that is
propagated by them. The freedom in $x_0(\bar\tau)$ is to be used to
make $D=1$ at $\bar x=1$. Now $\bar x=0$ and $\bar x=1$ resemble
``regular singular points'', if we are prepared to generalize this
concept from linear ODEs to nonlinear PDEs. Near $\bar x=0$, the four
evolution equations are clearly of the form $\partial Z/\partial \bar
x = {\rm regular}/\bar x$. That $\bar x=1$ is also a regular singular
point becomes clearest if we replace $D$ by $\bar D=(1-D)/(\bar x-1)$.
The ``evolution'' equation for $X_+$ near $\bar x=1$ then takes the
form $\partial X_+/\partial \bar x = {\rm regular}/(\bar x -1)$, while
the other three equations are regular.

This format of the equations also demonstrates how to restrict from a
DSS to a CSS ansatz: one simply drops the $\bar
\tau$-derivatives. The constraint then becomes algebraic, and the
resulting ODE system can be considered to have three rather than four
dependent variables. 

Given that the critical solutions are regular at the past branch of
$r=0$ and at the past sound cone of the singularity, and that they are
self-similar, one would expect them to be singular at the future light
cone of the singularity (because after solving the boundary value
problem there is no free parameter left in the solution). The real
situation is more subtle as we shall see in Section
\ref{section_singularity}.

As a final remark, it appears that all critical solutions found so far
for any matter model, of both type I and type II (see section
\ref{section_massgap} below), do not admit a limit $G\to 0$, so that
they are only brought into existence by gravity.


\subsection{Universality and scaling} \label{section_scaling}


We have seen that the universal solution arising in critical
collapse can be constructed semi-analytically from a self-similar
ansatz plus regularity conditions. The fact that it is universal up to
fine-tuning of one parameter is equivalent to its being an attractor
of codimension one. The linearization of that statement around the
critical solution is that it has exactly one unstable mode.

We now formulate this idea more precisely. For simplicity of notation,
we limit ourselves to the spherically symmetric CSS case, for example
the perfect fluid. The DSS case is discussed in \cite{Gundlach_Chop2}.
Let $Z$ stand for a set of scale-invariant variables of the problem in
a first-order formulation. $Z(r)$ is an element of the phase space,
and $Z(r,t)$ a solution. The self-similar solution is of the form
$Z(r,t)=Z_*(-r/t)=Z_*(x)$. [We have chosen the Schwarzschild-like
coordinates defined in Eqn. (\ref{tr_metric}), have shifted the origin
of $t$ to $t=t_*$, and consider only values $t<0$.]  In the echoing
region, where $Z_*$ dominates, we linearize around it. As the
background solution is $\tau$-independent, $Z(x,\tau)=Z_*(x)$, its
linear perturbations can depend on $\tau$ only exponentially (with
complex exponent $\lambda$), that is
\begin{equation}
\delta Z(x,\tau)=\sum_{i=1}^\infty C_i \, e^{\lambda_i \tau} f_i(x),
\end{equation}
where the $C_i$ are free constants.  We can also write this in the
more familiar space and time coordinates $r$ and $t$
\begin{equation}
r = Lxe^{-\tau}, \quad t = - Le^{-\tau}
\end{equation}
already defined in (\ref{x_tau}) above.  To linear order, the solution
in the echoing region is then of the form
\begin{equation}
Z(r,t) \simeq Z_*\left(-{r\over t}\right) + \sum_{i=1}^\infty C_i(p)
\left(-{t\over L}\right)^{-\lambda_i} f_i\left(-{r\over t}\right).
\end{equation} 
The coefficients $C_i$ depend in a complicated way on the initial
data, and hence on $p$. If $Z_*$ is a critical solution, by definition
there is exactly one $\lambda_i$ with positive real part (in fact it
is purely real), say $\lambda_1$. As $t\to 0^-$, all other perturbations
vanish. In the following we consider this limit, and retain only
the one growing perturbation. By definition the critical
solution corresponds to $p=p_*$, so we must have
$C_1(p_*)=0$. Linearizing around $p_*$, we obtain
\begin{equation}
\label{echoing_region}
\lim_{t\to 0} Z(r,t) \simeq Z_*\left(-{r\over t}\right) 
+ {dC_1\over dp} (p-p_*)
\left(-{t\over L}\right)^{-\lambda_1} f_1\left(-{r\over t}\right).
\end{equation}
This approximate solution explains why the solution $Z_*$ is
universal.  It is now also clear why Eqn. (\ref{duration}) holds, that
is why we see more of the universal solutions (in the DSS case, more
``echos'') as $p$ is tuned closer to $p_*$. At an intuitive level, the
picture is of either a limit point (in the CSS case), or limit cycle
(in the DSS case, as in Fig. 1), in phase space, which is in an
attractor in the hypersurface separating black hole from no black hole
data. We shall reconsider this picture below in section
\ref{section_RGflow}. The universal solution is also called the
critical solution because it would be revealed up to the singularity
$\tau=\infty$ if perfect fine-tuning of $p$ would be possible. A
possible source of confusion here is that the critical solution,
because it is self-similar, is not asymptotically flat. Nevertheless,
it can arise in a region up to finite radius as the limiting case of a
family of asymptotically flat solutions. At large radius, it is
matched to an asymptotically flat solution which is not universal but
depends on the initial data (as does the place of matching.)

The following calculation of the critical exponent from the linear
perturbations of the critical solution by dimensional analysis was
suggested by Evans and Coleman
\cite{EvansColeman} and carried out by Koike, Hara and Adachi
\cite{KoikeHaraAdachi} and Maison \cite{Maison}. It was generalized
to the discretely self-similar (DSS) case by Gundlach
\cite{Gundlach_Chop2}. For simplicity of notation 
we consider again the CSS case.

The solution has the approximate form (\ref{echoing_region}) over a
range of $t$. Now we extract Cauchy data at one particular value of
$t$ within that range, namely $t_p$ defined by
\begin{equation}
{dC_1\over dp} (p-p_*)
(-t_p)^{-\lambda_1} \equiv \epsilon,
\end{equation}
where $\epsilon$ is some constant $\ll 1$, so that at $t_p$ the linear
approximation is still valid. (The suffix $p$ indicates that $t_p$
depends on $p$.) At sufficiently small $-t$, the linear perturbation
has grown so much that the linear approximation breaks down. Later on
a black hole forms.  The crucial point is that we need not follow this
evolution in detail, nor does it matter at what amplitude $\epsilon$
we consider the perturbation as becoming non-linear. It is sufficient
to note that the Cauchy data at $t=t_p$ depend on $r$ only in the
combination $r/t_p$, namely
\begin{equation}
Z(r,t_p) \simeq Z_*\left(-{r\over t_p}\right) + \epsilon \
f_1\left(-{r\over t_p} \right).
\end{equation}
($t_p$ has of course been defined just so that the coefficient of
$f_1$ in this expression is the same for all values of $p$, namely
$\epsilon$.)  Furthermore the field equations do not have an intrinsic
scale. It follows that the solution based on those data must be {\it
exactly}
\cite{HE2} of the form
\begin{equation}
Z(r,t) = f\left({r\over t_p}, {t\over t_p}\right), 
\end{equation}
for some function $f$, throughout, even when the black hole forms and
perturbation theory breaks down, and later still after it has settled
down and the solution no longer depends on $t$. (This solution holds
only for $t>t_p$, because in its initial data we have neglected the
perturbation modes with $i>1$, which would be growing, not decaying,
towards the past.)  Because the black hole mass has dimension length,
it must be proportional to $t_p$, the only length scale in the
solution,
\begin{equation}
M \propto t_p \propto (p-p_*)^{1\over \lambda_1},
\end{equation}
and we have found the critical exponent $\gamma = 1/\lambda_1$. 

When the critical solution is DSS, the scaling law is modified.  This
was predicted by \cite{Gundlach_Chop2} and predicted independently and
verified in collapse simulations by Hod and Piran
\cite{HodPiran_wiggle}. On the straight line relating $\ln M$ to
$\ln(p-p_*)$, a periodic ``wiggle'' or ``fine structure'' of small
amplitude is superimposed:
\begin{equation}
\ln M = \gamma \ln (p-p_*) + c + f[\gamma \ln (p-p_*) + c],
\end{equation}
with $f(z)=f(z+\Delta)$.  The periodic function $f$ is again universal
with respect to families of initial data, and there is only one
parameter $c$ that depends on the family of initial data,
corresponding to a shift of the wiggly line in the $\ln(p-p_*)$
direction. (No separate adjustment in the $\ln M$ direction is
possible.)


\section{Extensions of the basic scenario}


\subsection{More matter models}


Choptuik's results have been confirmed for a variety of other matter
models. In some of these, qualitatively new phenomena were
discovered, and we review this body of work by phenomena rather than
chronologically or by matter models. A presentation by matter models is
given in Table 1 for completeness.

An exceptional case is spherically symmetric dust collapse. Here,
the entire spacetime, the Tolman-Bondi solution, is given in closed
form from the initial velocity and density profiles. Excluding shell
crossing singularities, there is a ``phase transition'' between
initial data forming naked singularities at the center and data
forming black holes. Which of the two happens depends only the leading
terms in an expansion of the initial data around $r=0$
\cite{Christodoulou0,Jhingan}. One could argue that this fact also
makes the matter model rather unphysical.

\begin{table}
\caption{An overview of numerical work in critical collapse. Question marks
denote missing links.}
\label{table_mattermodels}
\begin{tabular}{l | l | l | l}
\hline
Matter model & Collapse  & Critical &
Perturbations \\ 
& simulations & solution & \\
\hline
{\it Perfect fluid} &&& \\
-- $k=1/3$ & \cite{EvansColeman} & \cite{EvansColeman} &
\cite{KoikeHaraAdachi} \\
-- general $k$ & ?  & \cite{Maison} & \cite{Maison} \\
\hline
{\it Real scalar field} &&& \\
-- massless, min. coupled & \cite{Choptuik91,Choptuik92,Choptuik94} &
\cite{Gundlach_Chop1} & \cite{Gundlach_Chop2} \\
-- massive & \cite{Choptuik94,BradyChambersGoncalves} 
& \cite{HaraKoikeAdachi,GundlachMartin}\footnotemark[1] 
& \cite{HaraKoikeAdachi,GundlachMartin}\footnotemark[1] \\ 
-- conformally coupled &
\cite{Choptuik92} & ?  & ?  \\
\hline
{\it 2-d sigma model} &&& \\
-- complex scalar ($\kappa=0$) & \cite{Choptuik_pc} &
\cite{Gundlach_Chop2}\footnotemark[2], \cite{HE1}\footnotemark[3]  &
\cite{Gundlach_Chop2}\footnotemark[2], \cite{HE2}\footnotemark[3]\\
-- axion-dilaton ($\kappa=1$) & 
\cite{HamadeHorneStewart} & \cite{EardleyHirschmannHorne,HamadeHorneStewart} & 
\cite{HamadeHorneStewart} \\
-- scalar-Brans-Dicke ($\kappa>0$) & \cite{LieblingChoptuik} & & \\ --
general $\kappa$ including $\kappa<0$ & ? & \cite{HE3} & \cite{HE3} \\
\hline 
Scalar electrodynamics & \cite{HodPiran_charge} &
\cite{GundlachMartin}\footnotemark[4] &
\cite{GundlachMartin}\footnotemark[4]  \\ 
\hline
$SU(2)$ Yang-Mills & \cite{ChoptuikChmajBizon} & \cite{Gundlach_EYM}
& \cite{Gundlach_EYM} \\
\hline
$SU(2)$ Skyrme model & \cite{BizonChmaj} & \cite{BizonChmaj}
& \cite{BizonChmaj} \\
\hline
Axisymmetric vacuum & \cite{AbrahamsEvans,AbrahamsEvans2} & ? & ? \\
\end{tabular}
\end{table}

\footnotetext[1]{The critical solution and its perturbations for the
massive scalar field are asymptotic to those of the massless scalar.}

\footnotetext[2]{The (DSS) critical solution for the real massless scalar
field is also the critical solution for the complex scalar field. The
additional perturbations are all stable \cite{Gundlach_Chop2}.}

\footnotetext[3]{There is also a CSS solution \cite{HE1}, but it has
three unstable modes, not only one \cite{HE2}.}

\footnotetext[4]{The scalar electrodynamics critical solution is again
the real scalar field critical solution. Its perturbations are
those of the complex scalar field.}



\subsection{CSS and DSS critical solutions}


As we have seen, a critical solution is one that sits on the boundary
of black hole formation, and has exactly one ``growing mode'', so that
it acts as an intermediate attractor (Evans). All one-parameter
families of initial data crossing that boundary are then ``funnelled''
(Eardley) through that one solution. So far, we have seen an example
each of a critical solution with discrete and with continuous
self-similarity. There may be regular CSS or DSS solutions with more
than one growing mode, but they will not appear in Choptuik type
fine-tuning. An example for this is provided by the spherically
symmetric massless complex scalar field.  Hirschmann and Eardley
\cite{HE1} found a way of constructing a CSS
scalar field solution by making the scalar field $\phi$ complex but
limiting it to the ansatz
\begin{equation}
\phi=e^{i\omega\tau} f(x),
\end{equation}
with $\omega$ a real constant and $f$ real. The metric is then
homothetic, while the scalar field shows a trivial kind of ``echoing''
in the complex phase. Later, they found that this solution has three
modes with $Re\lambda>0$ \cite{HE2} and is therefore not the critical
solution. Gundlach \cite{Gundlach_Chop2} examined complex scalar field
perturbations around Choptuik's real scalar field critical solution
and found that only one of them, purely real, has $Re\lambda>0$, so
that the real scalar field critical solution is a critical solution
(up to an overall complex phase) also for the free complex scalar
field. This had been seen already in collapse calculations
\cite{Choptuik_pc}. 

As the symmetry of the critical solution, CSS or DSS, depends on the
matter model, it is interesting to investigate critical behavior in
parameterized families of matter models. Two such one-parameter
families have been investigated. The first one is the spherical
perfect fluid with equation of state $p=k\rho$ for arbitrary
$k$. Maison \cite{Maison} constructed the regular CSS solutions and
its linear perturbations for a large number of values of $k$. In each
case, he found exactly one growing mode, and was therefore able to
predict the critical exponent. (To my knowledge, these critical
exponents have not yet been verified in collapse simulations.) As Ori
and Piran before \cite{OriPiran}, he found that there are no regular
CSS solutions for $k> 0.88$. There is nothing in the equation of state
to explain this. In particular, the perfect fluid is well behaved up
to $k<1$. It remains unknown what happens in critical collapse for
$k>0.88$ Black hole formation may begin with a minimum mass. (In the
absence of a mass scale in the field equations, this mass gap would
depend on the family.) Alternatively, there may be a DSS critical
solution. The fact that the $k=1$ perfect fluid is equivalent to a
massless scalar field, which does have a DSS critical solution, hints
in this direction. Nevertheless, a scalar field solution corresponds
to a perfect fluid solution only if $\phi_{,a}$ is everywhere
timelike, and this is not true for Choptuik's universal solution.

The second one-parameter family of matter models was suggested by
Hirschmann and Eardley \cite{HE3}, who looked for a natural way of
introducing a non-linear self-interaction for the (complex) scalar
field without introducing a scale. (We discuss dimensionful coupling
constants in the following sections.) They investigated the model
described by the action
\begin{equation}
S=\int \sqrt{g}\left(R-{2|\nabla\phi|^2\over
(1-\kappa|\phi|^2)^2}\right).
\end{equation}
Note that $\phi$ is now complex, and the parameter $\kappa$ is real
and dimensionless. This is a 2-dimensional sigma model with a target
space metric of constant curvature (namely $\kappa$), minimally
coupled to gravity. Moreover, for $\kappa > 0$ there are (nontrivial)
field redefinitions which make this model equivalent to a real
massless scalar field minimally coupled to Brans-Dicke gravity, with
the Brans-Dicke coupling given by
\begin{equation}
\omega_{\rm BD}=-{3\over2}+{1\over 8\kappa}.
\end{equation}
In particular, $\kappa=1$ ($\omega_{\rm BD}=-11/8$) corresponds to an
axion-dilaton system arising in string theory
\cite{EardleyHirschmannHorne}. $\kappa=0$ is the free complex scalar
field coupled to Einstein gravity).  Hirschmann and Eardley calculated
a CSS solution and its perturbations and concluded that it is the
critical solution for $\kappa>0.0754$, but has three unstable modes
for $\kappa<0.0754$. For $\kappa<-0.28$, it acquires even more
unstable modes. The positions of the mode frequencies $\lambda$ in the
complex plane vary continuously with $\kappa$, and these are just
values of $\kappa$ where a complex conjugate pair of frequencies
crosses the real axis. The results of Hirschmann and Eardley confirm
and subsume collapse simulation results by Liebling and Choptuik
\cite{LieblingChoptuik} for the scalar-Brans-Dicke system, and
collapse and perturbative results on the axion-dilaton system by
Hamad\'e, Horne and Stewart \cite{HamadeHorneStewart}. Where the CSS
solution fails to be the critical solution, a DSS solution takes
over. In particular, for $\kappa=0$, the free complex scalar field,
the  critical solution is just the real scalar field DSS solution of
Choptuik.


\subsection{Black hole thresholds with a mass gap} \label{section_massgap}


The first models in which critical phenomena were observed did not
have any length scales in the field equations. Later, models were
examined which have one such scale. Collapse simulations were carried
out for the spherically symmetric $SU(2)$
Einstein-Yang-Mills system by Choptuik, Chmaj and Bizon
\cite{ChoptuikChmajBizon}. In fine-tuning one-parameter families of
data to the black-hole threshold, they found two different kinds of
critical phenomena, dominated by two different critical
solutions. Which kind of behavior arises appears to depend on the
qualitative shape of the initial data. In one kind of behavior, black
hole formation turns on at an infinitesimal mass with the familiar
power-law scaling, dominated by a DSS critical solution. In the other
kind, black hole formation turns on at a finite mass, and the critical
solution is now a static, asymptotically flat solution which had been
found before by Bartnik and McKinnon \cite{BartnikMcKinnon}. It was
also known before that this solution (the least massive one of a
discrete family) had exactly two unstable perturbation modes
\cite{LavrelashviliMaison}.  The ansatz of Choptuik, Chmaj and Bizon
further allowed for only one of these unstable modes, with one sign of
these leading to collapse and the other to dispersion of the
solution. The Bartnik-McKinnon solution is then a critical solution
within this ansatz, in the sense of being an attractor of codimension
one on the black hole threshold. Choptuik, Chmaj and Bizon labelled
the two kinds of critical behavior type II and type I respectively,
corresponding to a second- and a first-order phase transition. The
newly found, type I critical phenomena show a scaling law that is
mathematically similar to the black hole mass scaling observed in type
II critical phenomena.  Let $\partial/\partial t$ be the static
Killing vector of the critical solution. Then the perturbed critical
solution is of the form
\begin{equation}
\label{typeIintermediate}
Z(r,t) = Z_*(r) + {d C_1\over dp} (p-p_*) 
e^{\lambda_1 t} f_1(r) + \hbox{decaying modes}.
\end{equation}
This is similar to Eqn. (\ref{echoing_region}), but the growth of the
unstable mode is now exponential in $t$, not in $\ln t$. We again
define a time $t_p$ by
\begin{equation}
{d C_1\over dp} (p-p_*) e^{\lambda_1 t_p} \equiv \epsilon,
\end{equation}
but now the initial data at $t_p$ are
\begin{equation}
Z(r,t_p) \simeq Z_*\left(r\right) + \epsilon \
f_1\left(r \right),
\end{equation}
so that that the final black hole mass is independent of $p-p_*$. (It
is of the order of the mass of the static critical solution.)
The scaling is only apparent in the lifetime of the critical solution,
which we can take to be $t_p$. It is
\begin{equation}
t_p = - {1\over \lambda_1} \ln(p-p_*) + {\rm const.}
\end{equation}

The type I critical solution can also have a discrete symmetry, that
is, can be periodic in time instead of being static. This behavior was
found in collapse situations of the massive scalar field by Brady,
Chambers and Gon\c calves \cite{BradyChambersGoncalves}. Previously,
Seidel and Suen \cite{SeidelSuen} had constructed periodic,
asymptotically flat, spherically symmetric self-gravitating massive
scalar field solutions they called oscillating soliton stars. By
dimensional analysis, the scalar field mass $m$ sets an overall scale
of $1/m$ (in units $G=c=1$). For given $m$, Seidel and Suen found a
one-parameter family of such solutions with two branches. The more
compact solution for a given ADM mass is unstable, while the more
extended one is stable to spherical perturbations. Brady, Chambers and
Gon\c calves (BCG) report that the type I critical solutions they find
are from the unstable branch of the Seidel and Suen solutions.
Therefore we are seeing a one-parameter family of (type I) critical
solutions, rather than an isolated critical solution. BCG in fact
report that the black hole mass gap does depend on the initial data.
They find a small wiggle in the mass of the critical solution which is
periodic in $\ln(p-p_*)$, and which should have the same explanation
\cite{Gundlach_Chop2} 
as that found in the mass of the black hole in type II DSS critical
behavior. If type I or type II behavior is seen appears to depend
mainly on the ratio of the length scale of the initial data to the
length scale $1/m$.

One point in the results of BCG is worth expanding on. In the critical
phenomena that were first observed, with an isolated critical
solution, only one number's worth of information, namely the
separation $p-p_*$ of the initial data from the black hole threshold,
survives to the late stages of the time evolution. This is true for
both type I and type II critical phenomena. In type II phenomena,
$p-p_*$ determines the black hole mass, while in both type I and II it
also determines the lifetime of the critical solution (the number of
echos). Recall that our definition of a critical solution is one that
has exactly one unstable perturbation mode, with a black hole formed
for one sign of the unstable mode, but not for the other. This
definition does not exclude an $n$-dimensional family of critical
solutions. Each solution in the family would then have $n$ marginal
modes leading to neighboring critical solutions, as well as the one
unstable mode. $n+1$ numbers' worth of information would survive from
the initial data, and the mass gap in type I, or the critical exponent
for the black hole mass in type II, for example, would depend on the
initial data through $n$ parameters. In other words, universality
would exist in diminished form. The results of BCG are an example of a
one-parameter family of type I critical solutions. Recently, Brodbeck
et al. \cite{Brodbecketal} have shown, under the assumption of
linearization stability, that there is a one-parameter family of
stationary, rotating solutions beginning at the (spherically
symmetric) Bartnik-McKinnon solution. This could turn out to be a
second one-parameter family of type I critical solutions, provided
that the Bartnik-McKinnon solution does not have any unstable modes
outside spherical symmetry (which has not yet been investigated)
\cite{Rendallpc}.

Bizo\'n and Chmaj have studied type I critical collapse of an $SU(2)$
Skyrme model coupled to gravity, which in spherical symmetry with a
hedgehog ansatz is characterized by one field $F(r,t)$ and one
dimensionless coupling constant $\alpha$. Initial data $F(r)\sim
\tanh(r/p)$, $\dot F(r)=0$ surprisingly form black holes for both
large and small values of the parameter $p$, while for an
intermediate range of $p$ the endpoint is a stable static solution
called a skyrmion. (If $F$ was a scalar field, one would expect only
one critical point on this family.) The ultimate reason for this
behavior is the presence of a conserved integer ``baryon number'' in
the matter model. Both phase transitions along this one-parameter
family are dominated by a type I critical solution, that is a
different skyrmion which has one unstable mode. In particular, an
intermediate time regime of critical collapse evolutions agrees well
with an ansatz of the form (\ref{typeIintermediate}), where $Z_*$,
$f_1$ and $\lambda$ were obtained independently. It is interesting to
note that the type I critical solution is singular in the limit
$\alpha \to 0$, which is equivalent to $G \to 0$, because the known
type II critical solutions for any matter model also do not have a
weak gravity limit.

Apparently, type I critical phenomena can arise even without the
presence of a scale in the field equations. A family of exact
spherically symmetric, static, asymptotically flat solutions of vacuum
Brans-Dicke gravity given by van Putten was found by Choptuik,
Hirschmann and Liebling \cite{ChopHirschLieb} to sit at the black
hole-threshold and to have exactly one growing mode. This family has
two parameters, one of which is an arbitrary overall scale.


\subsection{Approximate self-similarity and universality classes}


As we have seen, the presence of a length scale in the field equations
can give rise to static (or oscillating) asymptotically flat critical
solutions and a mass gap at the black hole threshold. Depending on the
initial data, this scale can also become asymptotically irrelevant as
a self-similar solution reaches ever smaller spacetime scales. This
behavior was already noticed by Choptuik in the collapse of a massive
scalar field, or one with a potential term generally
\cite{Choptuik94} and confirmed by  Brady,
Chambers and Gon\c calves \cite{BradyChambersGoncalves}. It was also
seen in the spherically symmetric EYM system
\cite{ChoptuikChmajBizon}. In order to
capture the notion of an asymptotically self-similar solution, one may
set the arbitrary scale $L$ in the definition (\ref{x_tau}) of $\tau$
to the scale of the field equations, here $1/m$.  

Introducing suitable dimensionless first-order variables $Z$ (such as
$a$, $\alpha$, $\phi$, $r\phi_{,r}$ and $r\phi_{,t}$ for the
spherically symmetric scalar field), one can write the field
equations as a first order system
\begin{equation}
F\left(Z,Z_{,x},Z_{,\tau},e^{-\tau}\right)=0.
\end{equation}
Every appearance of $m$ gives rise to an appearance of $e^{-\tau}$. If
the field equations contain only positive integer powers of $m$, one
can make an ansatz for the critical solution of the form
\begin{equation}
\label{asymptotic_CSS}
Z_*(x,\tau) = \sum_{n=0}^\infty e^{-n\tau} Z_n(x)
\end{equation}
where each $Z_n(x)$ is calculated recursively from the preceding
ones. For large enough $\tau$ (on spacetime scales small enough, close
enough to the singularity), this infinite series is expected to
converge. A similar ansatz can be made for the linear perturbations of
$Z_*$, and solved again recursively. Fortunately, one can calculate
the leading order background term $Z_0$ on its own, and obtain the
exact echoing period $\Delta$ in the process (in the case of
DSS). Similarly, one can calculate the leading order perturbation term
on the basis of $Z_0$ alone, and obtain the exact value of the
critical exponent $\gamma$ in the process. This procedure was carried
out by Gundlach \cite{Gundlach_EYM} for the Einstein-Yang-Mills
system, and by Gundlach and Mart\'\i n-Garc\'\i a
\cite{GundlachMartin} for massless scalar electrodynamics. Both
systems have a single scale $1/e$ (in units $c=G=1$), where $e$ is the
gauge coupling constant. 

The leading order term $Z_0$ in the expansion of the self-similar
critical solution $Z_*$ obeys the equation
\begin{equation}
F\left(Z_0,Z_{0,x},Z_{0,\tau},0\right)=0.
\end{equation}
Clearly, the critical solution is independent of the overall scale
$L_0$. By a similar argument, so are its perturbations, and therefore
the critical exponent $\gamma$. Therefore, all systems with a single
length scale $L_0$ in the field equations are in one universality
class
\cite{HaraKoikeAdachi,GundlachMartin}. 
The massive scalar field, for any value of $m$, or massless scalar
electrodynamics, for any value of $e$, are in the same universality
class as the massless scalar field. This notion of universality
classes is fundamentally the same as in statistical mechanics.

If there are several scales $L_0$, $L_1$, $L_2$ etc. present in the
problem, a possible approach is to set the arbitrary scale in
(\ref{x_tau}) equal to one of them, say $L_0$, and define the
dimensionless constants $l_i=L_i/L_0$ from the others.  The size of
the universality classes depends on where the $l_i$ appear in the
field equations. If a particular $L_i$ appears in the field equations
only in positive integer powers, the corresponding $l_i$ appears only
multiplied by $e^{-\tau}$, and will be irrelevant in the scaling
limit. All values of this $l_i$ therefore belong to the same
universality class. As an example, adding a quartic self-interaction
$\lambda\phi^4$ to the massive scalar field, for example, gives rise
to the dimensionless number $\lambda/m^2$, but its value is an
irrelevant (in the language of renormalisation group theory)
parameter. All self-interacting scalar fields are in fact in the same
universality class. Contrary to the statement in
\cite{GundlachMartin}, I would now conjecture that massive scalar
electrodynamics, for any values of $e$ and $m$, forms a single
universality class in type II critical phenomena. Examples of
dimensionless parameters which do change the universality class are
the $k$ of the perfect fluid, the $\kappa$ of the 2-dimensional sigma
model, or a conformal coupling of the scalar field.


\subsection{Beyond spherical symmetry} \label{section_nonspherical}


Every aspect of the basic scenario: CSS and DSS, universality and
scaling applies directly to a critical solution that is not
spherically symmetric, but all the models we have described are
spherically symmetric. There are only two exceptions to date: a
numerical investigation of critical collapse in axisymmetric pure
gravity
\cite{AbrahamsEvans}, and a study of the nonspherical perturbations the
perfect fluid critical solution \cite{Gundlach_nonspherical}.  They
correspond to two related questions in going beyond spherical
symmetry.  Are there critical phenomena in gravitational collapse far
from spherical symmetry? And: are the critical phenomena in the known
spherically symmetric examples destroyed by small deviations from
spherical symmetry?


\subsubsection{Axisymmetric gravitational waves}


The paper of Abrahams and Evans \cite{AbrahamsEvans}
was the first paper on critical collapse to be published after
Choptuik's PRL, but it remains the only one to investigate a
non-spherically symmetric situation, and therefore also the only one
to investigate critical phenomena in the collapse of gravitational
waves in vacuum. Because of its importance, we summarize its
contents here with some technical detail.

The physical situation under consideration is axisymmetric vacuum
gravity. The numerical scheme uses a 3+1 split of the spacetime. The
ansatz for the spacetime metric is
\begin{equation}
\label{axisymmetric}
ds^2=-\alpha^2\,dt^2 + \phi^4\left[e^{2\eta/3}(dr+\beta^r\,dt)^2 +
r^2e^{2\eta/3}(d\theta+\beta^\theta\,dt)^2 +
e^{-4\eta/3}r^2\sin^2\theta\,d\varphi^2\right],
\end{equation}
parameterized by the lapse $\alpha$, shift components $\beta^r$ and
$\beta^\theta$, and two independent coefficients $\phi$ and $\eta$ in
the 3-metric. All are functions of $r$, $t$ and $\theta$. The fact
that $dr^2$ and $r^2\,d\theta^2$ are multiplied by the same
coefficient is called quasi-isotropic spatial gauge. The variables for
a first-order-in-time version of the Einstein equations are completed
by the three independent components of the extrinsic curvature,
$K^r_\theta$, $K^r_r$, and $K^\varphi_\varphi$.  In order to obtain
initial data obeying the constraints, $\eta$ and $K^r_\theta$ are
given as free data, while the remaining components of the initial
data, namely $\phi$, $K^r_r$, and $K^\varphi_\varphi$, are determined
by solving the Hamiltonian constraint and the two independent
components of the momentum constraint respectively.  There are five
initial data variables, and three gauge variables. Four of the five
initial data variables, namely $\eta$, $K^r_\theta$, $K^r_r$, and
$K^\varphi_\varphi$, are updated from one time step to the next via
evolution equations. As many variables as possible, namely $\phi$ and
the three gauge variables $\alpha$, $\beta^r$ and $\beta^\theta$, are
obtained at each new time step by solving elliptic equations. These
elliptic equations are the Hamiltonian constraint for $\phi$, the
gauge condition of maximal slicing (${K_i}^i=0$) for $\alpha$, and the
gauge conditions $g_{\theta\theta}=r^2 g_{rr}$ and $g_{r\theta}=0$ for
$\beta^r$ and $\beta^\theta$ (quasi-isotropic gauge).

For definiteness, the two free functions, $\eta$ and $K^r_\theta$, in
the initial data were chosen to have the same functional form they
would have in a linearized gravitational wave with pure $(l=2,m=0)$
angular dependence. Of course, depending on the overall amplitude of
$\eta$ and $K^r_\theta$, the other functions in the initial data will
deviate more or less from their linearized values, as the non-linear
initial value problem is solved exactly. In axisymmetry, only one of
the two degrees of freedom of gravitational waves exists. In order to
keep their numerical grid as small as possible, Abrahams and Evans
chose the pseudo-linear waves to be purely ingoing. (In nonlinear
general relativity, no exact notion of ingoing and outgoing waves
exists, but this ansatz means that the wave is initially ingoing in
the low-amplitude limit.) This ansatz (pseudo-linear, ingoing, $l=2$),
reduced the freedom in the initial data to one free function of
advanced time, $I^{(2)}(v)$. A suitably peaked function was chosen.

Limited numerical resolution (numerical grids are now two-dimensional,
not one-dimensional as in spherical symmetry) allowed Abrahams and
Evans to find black holes with masses only down to $0.2$ of the ADM
mass. Even this far from criticality, they found power-law scaling of
the black hole mass, with a critical exponent $\gamma\simeq
0.36$. Determining the black hole mass is not trivial, and was done
from the apparent horizon surface area, and the frequencies of the
lowest quasi-normal modes of the black hole.  There was tentative
evidence for scale echoing in the time evolution, with $\Delta\simeq
0.6$, with about three echos seen. This corresponds to a scale range
of about one order of magnitude. By a lucky coincidence, $\Delta$ is
much smaller than in all other examples, so that several echos could
be seen without adaptive mesh refinement. The paper states that the
function $\eta$ has the echoing property $\eta(e^\Delta r,e^\Delta
t)=\eta(r,t)$. If the spacetime is DSS in the sense defined above, the
same echoing property is expected to hold also for $\alpha$, $\phi$,
$\beta^r$ and $r^{-1}\beta^\theta$, as one sees by applying the
coordinate transformation (\ref{x_tau}) to (\ref{axisymmetric}).

In a subsequent paper \cite{AbrahamsEvans2}, universality of the
critical solution, echoing period and critical exponent was
demonstrated through the evolution of a second family of initial data,
one in which $\eta=0$ at the initial time. In this family, black hole
masses down to $0.06$ of the ADM mass were achieved.
Further work on critical collapse far away from spherical symmetry
would be desirable, but appears to be held up by numerical
difficulty. 


\subsubsection{Perturbing around sphericity}


A different, and technically simpler, approach is to take a known
critical solution in spherical symmetry, and perturb it using
nonspherical perturbations.  Addressing this perturbative question,
Gundlach
\cite{Gundlach_nonspherical} has studied the generic non-spherical
perturbations around the critical solution found by Evans and Coleman
\cite{EvansColeman} for the $p={1\over3}\rho$ perfect fluid in
spherical symmetry. He finds that there is exactly one spherical
perturbation mode that grows towards the singularity (confirming the
previous results
\cite{KoikeHaraAdachi,Maison}). He finds no growing nonspherical modes at
all.

The main significance of this result, even though it is only
perturbative, is to establish one critical solution that really has
only one unstable perturbation mode within the full phase space.  As
the critical solution itself has a naked singularity (see Section
\ref{section_singularity}), this means that there is, for this
matter model, a set of initial data of codimension one in the full
phase space of general relativity that forms a naked singularity.  In
hindsight, this result also fully justifies the attention that
critical phenomena gravitational collapse have won as a ``natural''
route to naked singularities.


\subsection{Critical phenomena and naked singularities}
\label{section_singularity}


Choptuik's result have an obvious bearing on the issue of cosmic
censorship. (For a general review of cosmic censorship, see
\cite{Wald_censorship}.) As we shall see in this section, the critical
spacetime has a naked singularity. This spacetime can be approximated
arbitrarily well up to fine-tuning of a generic parameter. A region of
arbitrarily high curvature is seen from infinity as fine-tuning is
improved. Critical collapse provides a set of smooth initial data from
which a naked singularity is formed. In spite of news to the contrary,
it violates neither the letter nor the spirit of cosmic censorship
because this set is of measure zero. Nevertheless it comes closer than
would have been imagined possible before the work of Choptuik. First
of all, the set of data is of codimension one, certainly in the space
of spherical asymptotically flat data, and apparently
\cite{Gundlach_nonspherical} also in the space of all asymptotically
flat data. This means that one can fine-tune any generic parameter,
whichever comes to hand, as long as it parameterizes a smooth curve in
the space of initial data. Secondly, critical phenomena seem to be
generic with respect to matter models, including realistic matter
models with intrinsic scales. For a hypothetical experiment to create
a Planck-sized black hole in the laboratory through a strong
explosion, this would mean that one could fine-tune any one design
parameter of the bomb to the black hole threshold, without requiring
much control over its detailed effects on the explosion.

The metric of the critical spacetime is of the form $e^{-2\tau}$ times
a regular metric. From this general form alone, one can conclude that
$\tau=\infty$ is a curvature singularity, where Riemann and Ricci
invariants blow up like $e^{4\tau}$, and which is at finite proper
time from regular points.  The Weyl tensor with index position
${C^a}_{bcd}$ is conformally invariant, so that components with this
index position remain finite as $\tau\to\infty$. In this property it
resembles the initial singularity in Penrose's Weyl tensor conjecture
rather than the final singularity in generic gravitational
collapse. This type of singularity is called ``conformally
compactifiable'' \cite{Tod_pc} or ``isotropic'' \cite{Goodeetal}. Is
the singularity naked, and is it timelike, null or a ``point''? The
answer to these questions remains confused, partly because of
coordinate complications, partly because of the difficulty of
investigating the singular behavior of solutions numerically.

Choptuik's, and Evans and Coleman's, numerical codes were limited to
the region $t<0$, in the Schwarzschild-like coordinates
(\ref{tr_metric}), with the origin of $t$ adjusted so that the
singularity is at $t=0$. Evans and Coleman conjectured that the
singularity is shrouded in an infinite redshift based on the
fact that $\alpha$ grows as a small power of $r$ at constant $t$. This
is directly related to the fact that $a$ goes to a constant
$a_\infty>1$ as $r\to\infty$ at constant $t$, as one can see from the
Einstein equation (\ref{dalpha_dr}). This in turn means simply that
the critical spacetime is not asymptotically flat, but asymptotically
conical at spacelike infinity, with the Hawking mass proportional to
$r$. Hamad\'e and Stewart
\cite{HamadeStewart} evolved near-critical scalar field spacetimes on
a double null grid, which allowed them to follow the time evolution up
to close to the future light cone of the singularity. They found
evidence that this light cone is not preceded by an apparent horizon,
that it is not itself a (null) curvature singularity, and that there
is only a finite redshift along outgoing null geodesics slightly
preceding it. (All spherically symmetric critical spacetimes appear to
be qualitatively alike as far as the singularity structure is
concerned, so that what we say about one is likely to hold for the
others.)

Hirschmann and Eardley \cite{HE1} were the first to continue a
critical solution itself right up to the future light cone. They
examined a CSS complex scalar field solution that they had constructed
as a nonlinear ODE boundary value problem, as discussed in Section
\ref{section_regular}. (This particular one is not a proper
critical solution, but that should not matter for the global
structure.) They continued the ODE evolution in the self-similar
coordinate $x$ through the coordinate singularity at $t=0$ up to the
future light cone by introducing a new self-similarity coordinate $x$.
The self-similar ansatz reduces the field equations to an ODE
system. The past and future light cones are regular singular points of
the system, at $x=x_1$ and $x=x_2$. At these ``points'' one of the two
independent solutions is regular and one singular. The boundary value
problem that originally defines the critical solution corresponds to
completely suppressing the singular solution at $x=x_1$ (the past
light cone). The solution can be continued through this point up to
$x=x_2$. There it is a mixture of the regular and the singular
solution.

We now state this more mathematically. The ansatz of Hirschmann and
Eardley for the self-similar complex scalar field is (we slightly
adapt their notation)
\begin{equation}
\phi(x,\tau) = f(x) e^{i\omega\tau}, \quad a=a(x), \quad
\alpha=\alpha(x),
\end{equation}
with $\omega$ a real constant. Near the future light cone they find
that $f$ is
approximately of the form
\begin{equation}
f(x)\simeq C_{\rm reg}(x) + (x-x_2)^{(i\omega+1)(1+\epsilon)} C_{\rm
sing}(x),
\end{equation}
with $C_{\rm reg}(x)$ and $C_{\rm sing(x)}$ regular at $x=x_2$, and
$\epsilon$ a small positive constant. The singular part of the scalar
field oscillates an infinite number of times as $x\to x_2$, but with
decaying amplitude. This means that the scalar field $\phi$ is just
differentiable, and that therefore the stress tensor is just
continuous. It is crucial that spacetime is not flat, or else
$\epsilon$ would vanish. For this in turn it is crucial that the
regular part $C_{\rm reg}$ of the solution does not vanish, as one
sees from the field equations.

The only other case in which the critical solution has been continued
up to the future light cone is Choptuik's real scalar field solution
\cite{Gundlach_Chop2}. Let $X_+$ and $X_-$ be the ingoing and outgoing
wave degrees of freedom respectively defined in (\ref{Xpm}). At the
future light cone $x=x_2$ the solution has the form
\begin{eqnarray}
X_-(x,\tau) & \simeq & f_-(x,\tau), \\
X_+(x,\tau) & \simeq & f_+(x,\tau) + (x-x_2)^\epsilon f_{\rm sing}(x,\tau
- C\ln x) ,
\end{eqnarray}
where $C$ is a positive real constant, $f_-$, $f_+$ and $f_{\rm sing}$
are regular real functions with period $\Delta$ in their second
argument, and $\epsilon$ is a small positive real constant. (We have
again simplified the original notation.) Again, the singular part of
the solution oscillates an infinite number of times but with decaying
amplitude. Gundlach concludes that the scalar field, the metric
coefficients, all their first derivatives, and the Riemann tensor
exist, but that is as far as differentiability goes. (Not all second
derivatives of the metric exist, but enough to construct the Riemann
tensor.) If either of the regular parts $f_-$ or $f_+$ vanished,
spacetime would be flat, $\epsilon$ would vanish, and the scalar field
itself would be singular. In this sense, gravity regularizes the
self-similar matter field ansatz. In the critical solution, it does
this perfectly at the past lightcone, but only partly at the future
lightcone. Perhaps significantly, spacetime is almost flat at the
future horizon in both the examples, in the sense that the Hawking
mass divided by $r$ is a very small number, as small as $10^{-6}$ (but
not zero according to numerical work by Horne
\cite{Horne_pc}) in the spacetime of Hirschmann and Eardley.

In summary, the future light cone (or Cauchy horizon) of these two
critical spacetimes is not a curvature singularity, but it is singular
in the sense that differentiability is lower than elsewhere in the
solution. Locally, one can continue the solution through the future
light cone to an almost flat spacetime (the solution is of course not
unique). It is not clear, however, if such a continuation can have a
regular center $r=0$ (for $t>0$), although this seems to have been
assumed by some authors. A priori, one should expect a conical
singularity, with a (small) defect angle at $r=0$.

The results just discussed were hampered by the fact that they are
investigations of singular spacetimes that are only known in numerical
form, with a limited precision. As an exact toy model we consider an
exact spherically symmetric, CSS solution for massless real scalar
field that was apparently first discovered by Roberts \cite{Roberts}
and then re-discovered in the context of critical collapse by Brady
\cite{Brady_Roberts} and Oshiro et al. \cite{Oshiro_Roberts}. We use
the notation of Oshiro et al. The solution can be given in double null
coordinates as
\begin{eqnarray}
ds^2 & = & - du\,dv + r^2(u,v) \,d\Omega^2, \\
r^2(u,v) & = & {1\over 4}\left[(1-p^2)v^2 - 2vu + u^2\right], \\
\phi(u,v) & = & {1\over 2} \ln {(1-p)v - u\over (1+p) v - u},
\end{eqnarray}
with $p$ a constant parameter. (Units $G=c=1$.) Two important
curvature indicators, the Ricci scalar and the Hawking mass, are
\begin{equation}
R={p^2 uv\over 2r^4}, \quad M = - {p^2 uv\over 8r}.
\end{equation}
The center $r=0$ has two branches, $u=(1+p)v$ in the past of $u=v=0$,
and $u=(1-p)v$ in the future.  For $0<p<1$ these are timelike
curvature singularities. The singularities have negative mass, and the
Hawking mass is negative in the past and future light cones. One can
cut these regions out and replace them by Minkowski space, not
smoothly of course, but without creating a $\delta$-function in the
stress-energy tensor. The resulting spacetime resembles the critical
spacetimes arising in gravitational collapse in some respects: it is
self-similar, has a regular center $r=0$ at the past of the curvature
singularity $u=v=0$ and is continuous at the past light
cone. It is also continuous at the future light cone, and the
future branch of $r=0$ is again regular. 

It is interesting to compare this with the genuine critical solutions
that arise as attractors in critical collapse. They are as regular as
the Roberts solution (analytic) at the past $r=0$, more regular
(analytic versus continuous) at the past light cone, as regular
(continuous) at the future light cone and, it is to be feared, less
regular at the future branch of $r=0$: In contrary to previous claims
\cite{HE1,Gundlach_Banach} there may be no
continuation through the future sound or light cone that does not have
a conical singularity at the future branch of $r=0$. The global
structure still needs to be clarified for all known critical
solutions.

In summary, the critical spacetimes that arise asymptotically in the
fine-tuning of gravitational collapse to the black-hole threshold have
a curvature singularity that is visible at infinity with a finite
redshift. The Cauchy horizon of the singularity is mildly singular
(low differentiability), but the curvature is finite there. It is
unclear at present if the singularity is timelike  or if
there exists a continuation beyond the Cauchy horizon with a regular
center, so that the singularity is limited, loosely speaking, to a
point. Further work should be able to clarify this. In any case, the
singularity is naked and the critical solutions therefore provide
counter-examples to some formulations of cosmic censorship which
state that naked singularities cannot arise from smooth initial data
in reasonable matter models. It is now clear that one must refine this
to state that there is no {\it open ball} of smooth initial for naked
singularities. Recent analytic work by Christodoulou also comes to
this conclusion \cite{Christodoulou5}.


\subsection{Black hole charge and angular momentum}


Given the scaling power law for the black hole mass in critical
collapse, one would like to know what happens if one takes a generic
one-parameter family of initial data with both electric charge and
angular momentum (for suitable matter), and fine-tunes the parameter
$p$ to the black hole threshold. Does the mass still show power-law
scaling? What happens to the dimensionless ratios $L/M^2$ and $Q/M$,
with $L$ the black hole angular momentum and $Q$ its electric charge?
Tentative answers to both questions have been given using perturbations around
spherically symmetric uncharged collapse.  


\subsubsection{Charge}


Gundlach and Mart\'\i n-Garc\'\i a
\cite{GundlachMartin} have studied scalar massless electrodynamics in
spherical symmetry. Clearly, the real scalar field critical solution
of Choptuik is a solution of this system too. Less obviously, it remains a
critical solution within massless (and in fact, massive) scalar
electrodynamics in the sense that it still has only one growing
perturbation mode within the enlarged solution space. Some of its
perturbations carry electric charge, but as they are all decaying,
electric charge is a subdominant effect. The charge of the black hole
in the critical limit is dominated by the most slowly decaying of the
charged modes. From this analysis, a universal power-law scaling of
the black hole charge
\begin{equation}
Q\sim (p-p_*)^\delta
\end{equation}
was predicted. The predicted value $\delta\simeq 0.88$ of the critical
exponent (in scalar electrodynamics) was subsequently verified in
collapse simulations by Hod and Piran
\cite{HodPiran_charge}. (The mass scales with $\gamma\simeq 0.37$ as
for the uncharged scalar field.) General considerations using
dimensional analysis led Gundlach and Mart\'\i n-Garc\'\i a to the
general prediction that the two critical exponents are always related,
for any matter model, by the inequality
\begin{equation}
\delta\ge2\gamma.
\end{equation}
This has not yet been verified in any other matter model. 


\subsubsection{Angular momentum}


Gundlach's results on non-spherically symmetric perturbations around
spherical critical collapse of a perfect fluid
\cite{Gundlach_nonspherical} allow for initial data, and therefore
black holes, with infinitesimal angular momentum.  All nonspherical
perturbations decrease towards the singularity. The situation is
therefore similar to scalar electrodynamics versus the real scalar
field. The critical solution of the more special model (here, the
strictly spherically symmetric fluid) is still a critical solution
within the more general model (a slightly nonspherical and slowly
rotating fluid). In particular, axial perturbations (also called
odd-parity perturbations) with angular dependence $l=1$ will determine
the angular momentum of the black hole produced in slightly
supercritical collapse. Using a perturbation analysis similar to that
of Gundlach and Mart\'\i n-Garc\'\i a \cite{GundlachMartin}, Gundlach
\cite{Gundlach_angmom} has derived the angular momentum scaling
\begin{equation}
\label{CSS_L}
\vec L = {\rm Re}\left[ (\vec A + i \vec B)
(p-p_*)^{\mu+i\omega} \right],
\end{equation}
where $\vec A$ and $\vec B$ are family-dependent constants, and the
complex critical exponent $\mu+i\omega$ is universal. For $p=\rho/3$,
he predicts the values of $\mu$ and $\omega$. In the special of
axisymmetry, this result reduces to
\begin{equation}
\label{axisymmL}
L=L_z=(p-p_*)^\mu A\cos[\omega\ln(p-p_*)+c],
\end{equation}
which is rather surprising. The explanation is of course that near the
black hole threshold the initial data, including the initial angular
momentum, are totally forgotten, while the oscillating angular
momentum of the black hole is a subdominant effect. These results have
not yet been tested against numerical collapse simulations.

Traschen \cite{Traschen} has drawn attention to a different possible
connection between critical phenomena and black hole charge. Consider
the equation of motion for a massive charged scalar test field on a
fixed charged black hole background. (The test field is coupled via
$\nabla + ieA$, but its back-reaction on both the metric and the
Maxwell field is neglected.) In the limit $Q=M$ for the background
black hole, and near the horizon this linear equation has a
scale-invariance not present in non-extremal black holes because the
surface gravity, which otherwise sets a scale, vanishes. The equation
therefore admits self-similar solutions. Traschen suggests these are
DSS, but there is no argument why either a CSS or a DSS solution
should play a special role.


\section{More speculative aspects}


\subsection{The renormalisation group as a time
evolution} \label{section_RGflow}


It has been pointed out by Argyres \cite{Argyres}, Koike, Hara and Adachi
\cite{KoikeHaraAdachi} and others that the time evolution near the
critical solution can be considered as a renormalisation group flow on
the space of initial data. The calculation of the critical exponent in
section \ref{section_scaling} is in fact mathematically identical with
that of the critical exponent governing the correlation length near
the critical point in statistical mechanics
\cite{Yeomans}, if one identifies the time evolution in the time
coordinate $\tau$ and spatial coordinate $x$ with the renormalisation
group flow.

For simple parabolic or hyperbolic differential equations, a discrete
renormalisation (semi)group acting on their solutions has been defined
in the following way \cite{Goldenfeld}. Evolve initial data over a
certain finite time interval, then rescale the final data in a certain
way. Solutions which are fixed points under this transformation are
scale-invariant, and may be attractors. In the context of the
spherically symmetric scalar field described in section
\ref{Choptuik_results} this prescription takes the following
form. Take free data $\phi_0(r)=\phi(t_0,r)$,
$\Pi_0(r)=\Pi(t_0,r)$. Evolve them from time $t_0$ to time
$t_1=e^{-\Delta} (t_0 - t_*) + t_*$.  Obtain new data $\phi_1(r) =
\phi(t_1,e^{-\Delta}r)$ and $\Pi_1(r) =
e^{-\Delta}\Pi(t_1,e^{-\Delta}r)$. One can introduce new coordinates
and fields such that the renormalisation transformation becomes a
simple time evolution without any explicit rescaling. For the scalar
field model, this form of the transformation is simply $Z_0(x) =
Z(\tau_0,x)$ to $Z_1(x) = Z(\tau_0+\Delta,x)$, where $Z$ stands for
the fields $\phi$ and $r\Pi$. The coordinates $x$ and $\tau$ replace
$r$ and $t$.

While this approach looks promising, its application to general
relativity has not yet been achieved. In general relativity as in
other field theories or in dynamical systems, a solution is determined
(at least locally) by an initial data set.  (In general relativity, a
solution is a spacetime, and the initial data are the first and second
fundamental forms of a spacelike hypersurface, plus suitable matter
variables.) A crucial distinctive feature of general relativity,
however, is that a solution does not correspond to a unique trajectory
in the space of initial data. This is because a spacetime can be
sliced in different ways, and on each slice one can have different
coordinate systems. Infinitesimally, this slicing and coordinate
freedom is parameterized by the lapse and shift. They can be set
freely, independently of the initial data, and they influence only the
coordinates on the spacetime, not the spacetime itself.

What coordinates should one use then when describing a time evolution
in GR as a renormalisation group transformation on the space of
initial data? For a given self-similar spacetime, there are preferred
coordinates adapted to the symmetry and defined by Eqn.
(\ref{DSS_coordinates}). These are far from unique. Nevertheless, one
can choose one of them, and then extend this choice of coordinates to
linear perturbations around the self-similar solution. In this linear
regime, one really obtains a time evolution such that a phase space
diagram of the type of Fig. 1 makes sense: On the limit cycle, the
same Cauchy data, expressed in the same space coordinates, return
periodically.  (The overall scale decreases in each period, and is
suppressed in Fig. 1.) Hara, Koike and Adachi
\cite{HaraKoikeAdachi} have used these coordinates 
in numerical work in order to find the spectrum of perturbations of
the critical solution by evolving a generic perturbation in time, and
peeling off the individual modes in the order of decreasing growth
rate (``Lyapunov analysis'').

Far away from the critical solution, no preferred gauge choice is
known. We are then faced with the general question: Given initial data
in general relativity, is there a prescription for the lapse and
shift, such that, if these are in fact data for a self-similar
solution, the resulting time evolution actively drives the metric to
the special form (\ref{DSS_coordinates}) that explicitly displays the
self-similarity? If such a prescription existed, one could try to find
the non-linear critical solution itself as a fixed point of a
renormalisation group transformation, as described by Bricmont and
Kupiainen \cite{BricmontKupiainen} for simple PDEs.  Dolan's
\cite{Dolan} description of a RG flow as a Hamiltonian flow may be
useful in making the identification.

An incomplete answer to this question has been provided by Garfinkle
\cite{Garfinkle}. His gauge conditions are 
\begin{equation}
\label{Garfinkle_gauge}
{\partial N\over \partial t} = {1\over3} N^2 K, \quad N^i=0,
\end{equation}
where $N$ is the lapse, $N^i$ the shift, and $K$ the trace of the
extrinsic curvature.  Garfinkle now
introduces the ``scale-invariant'' variables
\begin{equation}
\label{Garfinkle_scaleinvariant}
\tilde h_{ik} = N^2 h_{ik}, \quad \tilde K_{ik} = N^{-1} (K_{ik} - K
h_{ik}), \quad \omega_i = (\ln N)_{,i}.
\end{equation} 
$N$ will turn out to absorb the overall spacetime scale. The
introduction of $\omega_i$ has the purpose of extracting the
scale-invariant information from $N$.  Garfinkle gives an autonomous
system of equations for these degrees of freedom ($\tilde K_{ik}$ is
traceless) alone. The trace $K$ missing from this set is obtained
by solving the Hamiltonian constraint for $NK$.  The remaining degree
of freedom, $N$, is evolved in time by
Eqn. (\ref{Garfinkle_gauge}), but is not an active part of the
autonomous system.  If $NK$ is periodic in $t$, we can write the
spacetime metric in the form
\begin{equation}
ds^2 = e^{2Ct} \left(-\tilde N^2\, dt^2 + \tilde h_{ik}\, dx^i\,
dx^k\right),
\end{equation}
where the constant $C={1\over3}\langle NK\rangle$, while $\tilde N
\sim \exp {1\over3} \int (NK - \langle NK\rangle)\,dt$, with
$\langle\rangle$ the $t$-average. The constant $C$ can be set equal to
1 by a rescaling of the coordinate $t$. In the notation of this
review, $t$ is then the scale coordinate $\tau$.

The Einstein equations have now been split into an autonomous
scale-invariant part ($\tilde h_{ik},\tilde K_{ik}$) and a scale
variable ($N$) which is driven passively by the scale-invariant
part. When the scale-invariant variables are periodic in $t$
(independent of $t$), the spacetime is discretely (periodically)
self-similar. It seems unlikely, however, that the converse holds: If
we begin with data taken on an arbitrary hypersurface in a
self-similar spacetime, Garfinkle's evolution will not make his
scale-invariant variables periodic in $t$. This should happen only if
the initial data have been collected on the ``right'' initial
hypersurface. What is required is a gauge prescription that actively
drives the time evolution towards a slicing that makes the
scale-invariant variables periodic. 

Fixing the shift to be zero also poses a problem. In spherically
symmetric homothetic spacetimes the homothetic vector becomes
spacelike (inward pointing) at large radius on any spacelike slice. In
Garfinkle's ansatz the homothetic vector, if one exists, is identified
with $\partial/\partial t$, but with zero shift $\partial/\partial t$
is orthogonal to the hypersurfaces and therefore timelike, unlike the
homothetic vector in the known examples of critical spacetimes. It
seems necessary to allow for and prescribe a shift if
Garfinkle's scheme is to be consistent. In this context it may be
relevant that in the split between the scale-invariant and
scale part of the Einstein equations, the latter accounts only for one
degree of freedom. This seems to indicate that any such split is tied
to a slicing of spacetime.

Garfinkle suggests that a dynamical explanation of critical phenomena
would consist in finding a mechanism of ``dissipation'' that drives
the scale-invariant system towards a fixed point or a limit cycle. One
should keep in mind, however, that the limit cycle has at least one
unstable mode, or else naked singularities would be generic.


\subsection{Analytic approaches} 


A number of authors have attempted to explain critical collapse with
the help of analytic solutions. The one-parameter family of exact
self-similar real massless scalar field solutions first discovered by
Roberts \cite{Roberts} has already been presented in section
\ref{section_singularity}. It has been discussed in the context of
critical collapse by Brady
\cite{Brady_Roberts} and Oshiro et al. \cite{Oshiro_Roberts}, and
later Wang and Oliveira \cite{WangOliveira} and Burko
\cite{Burko}. The original, analytic, Roberts solution is cut and pasted to obtain a new solution
which has a regular center $r=0$ and which is asymptotically
flat. Solutions from this family with $p>1$ can be considered as black
holes, and to leading order around the critical value $p=1$, their
mass is $M\sim(p-p_*)^{1/2}$. The pitfall in this approach is that
only perturbations within the self-similar family are considered. But
the $p=1$ solution has many growing perturbations which are
spherically symmetric (but not self-similar), and is therefore not a
critical solution. This was already clear from the collapse
simulations at the black hole threshold, but recently Frolov
\cite{Frolov} has taken the trouble to calculate its perturbation
spectrum explicitly (within spherical symmetry). The perturbations can
be calculated analytically, and their spectrum is continuous, filling
a sector of the complex plane, with $Re\lambda\le1$. Soda and Hirata
\cite{SodaHirata} generalize the Roberts solution to higher spacetime
dimensions and calculate formal critical exponents. Oliveira and
Cheb-Terrab \cite{OliveiraCheb} generalize the Roberts solution to the
conformally coupled scalar field.

Also in the context of critical phenomena in gravitational collapse,
Koike and Mishima \cite{KoikeMishima} consider a two-parameter family
of solutions with a thin shell and an
outgoing null fluid, and formally derive a ``critical exponent'', but
there is no indication that these are critical solutions in the sense
of having exactly one unstable mode. Husain \cite{Husain} and Husain,
Martinez and N\'u\~nez \cite{HusainMartinezNunez2} find perfect fluid
``solutions'' by giving a metric and reading off the resulting
stress-energy tensor. The same authors \cite{HusainMartinezNunez1}
consider another scalar field exact solution (spherically symmetric
with a conformal Killing vector, but not homothetic) and obtain a
formal critical exponent of $1/2$. 

Peleg and Steif \cite{PelegSteif} have analyzed the collapse of thin
dust shells in 2+1 dimensional gravity with and without a cosmological
constant. There is a critical value of the shell's mass as a function
of its radius and position. The black hole mass scales with a critical
exponent of $1/2$. By analogy with the Roberts solution, it is likely
that in extending this mini-superspace model to a more general one, it
would reveal itself not be an attractor of codimension one.

Chiba and Soda \cite{ChibaSoda} have noticed that a conformal
transformation transforms Brans-Dicke gravity without matter into
general relativity with a massless, minimally coupled scalar
field. They transform the Choptuik solution from this so-called
Einstein frame back to the physical, or Jordan, frame, and obtain a
critical exponent for the formation of black holes in Brans-Dicke
gravity that depends on the Brans-Dicke coupling parameter. The
physical significance of this is doubtful in the absence of matter, as
$\omega$ is defined in the Einstein frame only through the coupling of
gravity to matter. [In the Einstein frame, all matter is minimally
coupled to the Einstein metric times the conformal factor
$\exp(\omega+{3\over2})^{1/2}\phi$.] Oliveira \cite{Oliveira} transforms
not the Choptuik but the Roberts solution.

Other authors have employed analytic approximations to the actual
Choptuik solution.  Pullin \cite{Pullin_Chop} has suggested describing
critical collapse approximately as a perturbation of the Schwarzschild
spacetime. Price and Pullin
\cite{PricePullin} have approximated the Choptuik solution by two flat
space solutions of the scalar wave equation that are matched at a
``transition edge'' at constant self-similarity coordinate $x$. The
nonlinearity of the gravitational field comes in through the matching
procedure, and its details are claimed to provide an estimate of the
echoing period $\Delta$. While the insights of this paper are
qualitative, some of its ideas reappear in the construction
\cite{Gundlach_Chop1} of the Choptuik solution as a 1+1 dimensional
boundary value problem.

In summary, purely analytic approaches have remained surprisingly
unsuccessful in dealing with critical collapse.


\subsection{Astrophysical applications?}


Any real world application of critical phenomena would require that
critical phenomena are not an artifact of the simple matter models
that have been studied so far, and that they are not an artifact of
spherical symmetry. At present this seems a reasonable
hypothesis. Critical collapse still requires a kind of fine-tuning of
initial data that does not seem to arise naturally in the
astrophysical world. Niemeyer and Jedamzik \cite{NiemeyerJedamzik} have
suggested a scenario that gives rise to such fine-tuning. In the early
universe, quantum fluctuations of the metric and matter can be
important, for example providing the seeds of galaxy formation. If
they are large enough, these fluctuations may even collapse
immediately, giving rise to what is called primordial black
holes. Large quantum fluctuations are exponentially more unlikely than
small ones, $P(\delta)\sim
\exp-\delta^2$, where $\delta$ is the density contrast of
the fluctuation. One would therefore expect the spectrum of primordial
black holes to be sharply peaked at the minimal $\delta$ that leads to
black hole formation. That is the required fine-tuning. In the
presence of fine-tuning, the black hole mass is much smaller
than the initial mass of the collapsing object, here the density
fluctuation. In consequence, the peak of the primordial black hole
spectrum might be expected to be at exponentially smaller values of
the black hole mass than expected naively.


\subsection{Critical collapse in semiclassical gravity}


As we have seen in the last section, critical phenomena may provide a
natural route from everyday scale down to much smaller scales, perhaps
down to the Planck scale. Various authors have investigated the
relationship of Choptuik's critical phenomena to quantum black
holes. It is widely believed that black holes should emit thermal
quantum radiation, from considerations of quantum field theory on a
fixed Schwarzschild background on the one hand, and from the purely
classical three laws of black hole mechanics on the other (see
\cite{Wald_BH} for a review). But there is no complete model of the
back-reaction of the radiation on the black hole, which should be
shrinking. In particular, it is unknown what happens at the endpoint
of evaporation, when full quantum gravity should become important. It
is debated in particular if the information that has fallen into the
black hole is eventually recovered in the evaporation process or lost.

To study these issues, various 2-dimensional toy models of gravity
coupled to scalar field matter have been suggested which are more or
less directly linked to a spherically symmetric 4-dimensional situation (see
\cite{Giddings_BH} for a review). In two space-time dimensions, the
quantum expectation value of the matter stress tensor can be
determined from the trace anomaly alone, together with the reasonable
requirement that the quantum stress tensor is conserved. Furthermore,
quantizing the matter scalar field(s) $f$ but leaving the metric
classical can be formally justified in the limit of many such matter
fields. The two-dimensional gravity used is not the two-dimensional
version of Einstein gravity but of a scalar-tensor theory of
gravity. $e^\phi$, where $\phi$ is called the dilaton, in the
2-dimensional toy model plays essentially the role of $r$ in 4
spacetime dimensions. There seems to be no preferred 2-dimensional toy
model, with arbitrariness both in the quantum stress tensor and in the
choice of the classical part of the model. In order to obtain a
resemblance of spherical symmetry, a reflecting boundary condition is
imposed at a timelike curve in the 2-dimensional spacetime. This plays
the role of the curve $r=0$ in a 2-dimensional reduction of the
spherically symmetric 4-dimensional theory.

How does one naively expect a model of semiclassical gravity to behave
when the initial data are fine-tuned to the black hole threshold?
First of all, until the fine-tuning is taken so far that curvatures on
the Planck scale are reached during the time evolution, universality
and scaling should persist, simply because the theory must approximate
classical general relativity. Approaching the Planck scale from above,
one would expect to be able to write down a critical solution that is
the classical critical solution asymptotically at large scales,
through an ansatz of the form
\begin{equation}
Z_*(x,\tau) = \sum_{n=0}^\infty e^{n\tau} Z_n(x),
\end{equation}
where the scale $L$ in $\tau=-\ln(-t/L)$ is now the Planck length.
This ansatz would recursively solve a semiclassical field equation,
where powers of $e^{\tau}$ (in coordinates $x$ and $\tau$) signal the
appearances of quantum terms.  Note that this is exactly the ansatz
(\ref{asymptotic_CSS}), but with the opposite sign in the exponent, so
that the higher order terms now become negligible as $\tau\to-\infty$,
that is away from the singularity on large scales. On the Planck scale
itself, this ansatz would not converge, and self-similarity would
break down.

Addressing the question from the side of classical general relativity,
Chiba and Siino \cite{ChibaSiino} write down their own 2-dimensional
toy model, and add a quantum stress tensor that is determined by the
trace anomaly and stress-energy conservation. They note that the
quantum stress tensor diverges at $r=0$. This means that the
additional quantum terms in the field equations carry powers not only
of $e^\tau$, but instead of $r^{-1}=x^{-1}e^{\tau}$. Hence 
no self-similar ansatz can be regular at the center $r=0$ ($x=0$) even
before the singularity appears at $\tau=\infty$. They conclude that
quantum gravity effects preclude critical phenomena on all scales,
even far from the Planck scale. More plausibly, their result indicates
that this 2-dimensional toy model does not capture essential physics.

At this point, Ayal and Piran \cite{AyalPiran} make an ad-hoc
modification to the semiclassical equations. They modify the quantum
stress tensor by a function which interpolates between 1 at large $r$,
and $r^2/L_p^2$ at small $r$. The stress tensor would only be
conserved if this function was a constant. The authors justify this
modification by pointing out that violation of energy conservation
takes place only at the Planck scale. It takes place, however, not
only where the solution varies dynamically on the Planck scale, but at
all times in a Planck-sized world tube around the center $r=0$, even
before the solution itself reaches the Planck scale dynamically. This
introduces a non-geometric, background structure, effect at the
world-line $r=0$. With this modification, Ayal and Piran obtain
results in agreement with our expectations set out above. For far
supercritical initial data, black formation and subsequent evaporation
are observed. With fine-tuning, as long as the solution stays away
from the Planck scale, critical solution phenomena including the
Choptuik universal solution and critical exponent are observed. (The
exponent is measured as $0.409$, indicating a limited accuracy of the
numerical method.) In an intermediate regime, the quantum effects
increase the critical value of the parameters $p$. This is interpreted
as the initial data partly evaporating while they are trying to form a
black hole.

Researchers coming from the quantum field theory side seem to favor a
model (the RST model) in which ad hoc ``counter terms'' have been
added to make it soluble. The matter is a conformally rather than
minimally coupled scalar field. The field equations are trivial up to
an ODE for a timelike curve on which reflecting boundary conditions
are imposed. The world line of this ``moving mirror'' is not clearly
related to $r$ in a 4-dimensional spherically symmetric model, but seems to
correspond to a finite $r$ rather than $r=0$. This may explain why the
problem of a diverging quantum stress tensor is not
encountered. Strominger and Thorlacius \cite{StromingerThorlacius}
find a critical exponent of $1/2$, but their 2-dimensional situation
differs from the 4-dimensional one in many aspects. Classically
(without quantum terms) any ingoing matter pulse, however weak, forms
a black hole. With the quantum terms, matter must be thrown in
sufficiently rapidly to counteract evaporation in order to form a
black hole. The initial data to be fine-tuned are replaced by
the infalling energy flux. There is a threshold value of the energy
flux for black hole formation, which is known in closed form. (Recall
this is a soluble system.) The mass of the black hole is defined as
the total energy it absorbs during its lifetime.  This black hole mass
is given by
\begin{equation}
M\simeq \left({\delta\over\alpha}\right)^{1\over2}
\end{equation}
where $\delta$ is the difference between the peak value of the flux
and the threshold value, and $\alpha$ is the quadratic order
coefficient in a Taylor expansion in advanced time of the flux around
its peak. There is universality with respect to different shapes of
the infalling flux in the sense that only the zeroth and second order
Taylor coefficients matter.

Peleg, Bose and Parker \cite{PelegBoseParker} study
the so-called CGHS 2-dimensional model. This (non-soluble) model does
allow for a study of critical phenomena with quantum effects turned
off. Again, numerical work is limited to integrating an ODE for the
mirror world line. Numerically, the authors find black hole mass
scaling with a critical exponent of $\gamma\simeq 0.53$. They find the
critical solution and the critical solution to be universal with
respect to families of initial data. Turning on quantum effects, the scaling
persists to a point, but the curve of $\ln M$ versus $\ln(p-p_*)$ then
turns smoothly over to a horizontal line. Surprisingly, the value of
the mass gap is not universal but depends on the family of initial
data. While this is the most ``satisfactory'' result among those
discussed here from the classical point of view, one should keep in
mind that all these results are based on mere toy models of quantum
gravity. 


\section{Summary and conclusions}


When one fine-tunes a one-parameter family of initial data to get
close enough to the black hole threshold, the details of the initial
data are completely forgotten in a spacetime region, and all
near-critical time evolutions look the same there. The only
information remembered from the initial data is how close one is to
the threshold. Either there is a mass gap (type I behavior), or black
hole formation starts at infinitesimal mass (type II behavior). In
type I, the universal critical solution is time-independent, or
periodic in time, and the better the fine-tuning, the longer it
persists. In type II, the universal critical solution is
scale-invariant or scale-periodic, and the better the fine-tuning, the
smaller the black hole mass, according to the famous formula
Eqn. (\ref{power_law}).

Both types of behavior arise because there is a solution which sits at
the black hole-threshold, and which is an intermediate attractor. The
basin of attraction is (at least locally) the black hole threshold
itself, pictured as a hypersurface of codimension one that bisects
phase space. Only the one perturbation mode pointing out of that
surface is unstable. Depending on its sign, the solution tips over
towards forming a black hole or towards dispersion. In the words of
Eardley, all one-parameter families of data trying to cross the black
hole threshold are funneled through a single time evolution.  If the
critical solution is time-independent, its linear perturbations grow
or decrease exponentially in time. If it is scale-invariant, they grow
or decrease exponentially with the logarithm of scale.  The power-law
scaling of the black hole mass follows by a clever application of
dimensional analysis. Although the mathematical foundations of this
approach remain doubtful at present, it allows precise numerical
calculation of the critical solution and the critical exponent, with
good agreement with numerical ``experiment''.

The importance of type II behavior lies in providing a natural route
from large to very small scales, with possible applications to
astrophysics and quantum gravity. Natural here means that the
phenomena persist for many simple matter models, without
counterexample so far, and, apparently, even beyond spherical
symmetry. As far as any generic parameter in the initial data provides
some handle on the amplitude of the one unstable mode, fine-tuning any
one generic parameter creates the phenomena. Moreover, scaling and
echoing are seen already quite far from the threshold in practice.

Clearly, more numerical work will be useful to further establish the
generality of the mechanism, or to find a counter-example instead. In
particular, future research should include highly non-spherical
situations, initial data with angular momentum and electric charge,
and matter models with a large number of internal degrees of freedom
(for example, collisionless matter instead of a perfect fluid). Going
beyond spherical symmetry, or including collisionless matter, will
certainly pose formidable numerical challenges.

The fundamental mathematical question in the field is why so many
matter models (in fact, all models investigated to date) admit a
critical solution, that is, an attractor of codimension one at the
black hole threshold. If the existence of a critical solution is
really a generic feature, then there should be at least an intuitive
argument, and perhaps a mathematical proof, for this important fact.
While it is at present unclear if all matter models admit a
self-similar solution with exactly one unstable mode(a type II
critical solution), no reasonable matter model can admit a
self-similar solution with no unstable modes, or naked singularities
would be endemic in nature. Again one wonders if there is a general
argument or proof for this fact.  Progress in understanding general
relativity as a dynamical system (in the presence of the slicing
freedom) may be a crucial step on the way towards these proofs, and
might also contribute to the study of singularities in general
relativity.

In the future, we can expect new phenomena based on continuous
families of critical solutions. In numerical investigations we may
also come across solutions on the black hole threshold with two or
more unstable modes, although as a matter of terminology I would not
call these ``critical'' solutions, because they would not
``naturally'' arise in collapse simulations.  Critical solutions so
far were scale-periodic or scale-invariant (type II), or static or
periodic in time (type I). Are solutions conceivable which have
neither of these symmetries, but which are still critical solutions in
the essential sense of being attractors inside the black hole threshold?

Numerical relativity has opened up a new research field in classical
general relativity, critical phenomena in gravitational collapse. The
interplay between numerical and analytic work in this new field is
still continuing strongly. While its surprising features have captured
the attention of many researchers in the GR community, it has also
thrown some light on the outstanding problem of mathematical
relativity, cosmic censorship.

I would like to thank A. Abrahams, P. Bizo\'n, B. Carr, A. Coley,
S. Jhingan and particularly A. Rendall for helpful comments on the
draft paper.



\begin{figure}
\label{fig_phasespace}
\epsfysize=10cm
\centerline{\epsffile{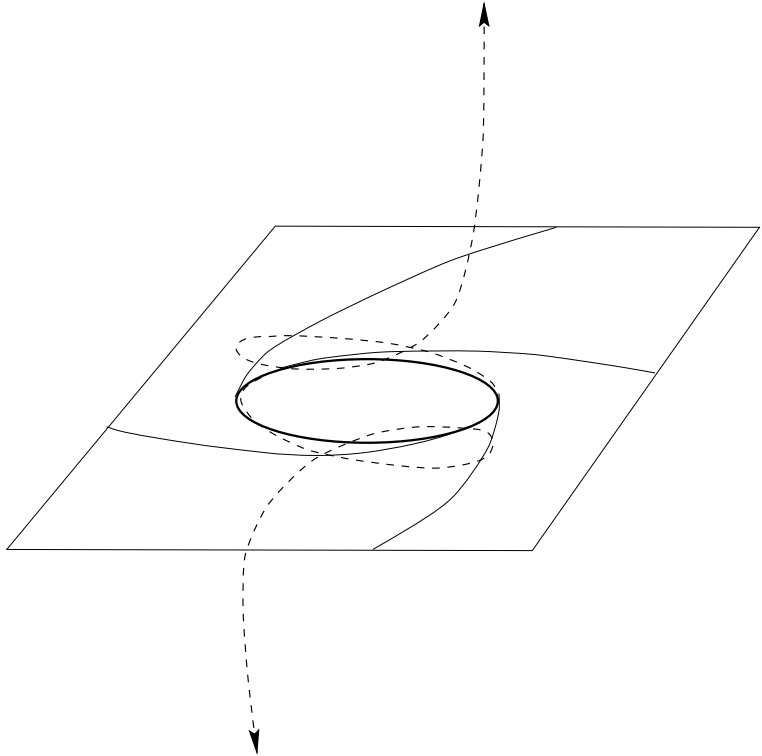}}
\caption{The phase space picture for discrete
self-similarity. The plane represents the critical surface. (In
reality this is a hypersurface of co-dimension one in an
infinite-dimensional space.) The circle (fat unbroken line) is the
limit cycle representing the critical solution. The thin unbroken
curves are spacetimes attracted to it. The dashed curves are
spacetimes repelled from it. There are two families of such curves,
labeled by one periodic parameter, one forming a black hole, the other
dispersing to infinity. Only one member of each family is shown.}
\end{figure}



\begin{figure}
\epsfysize=6cm
\centerline{\epsffile{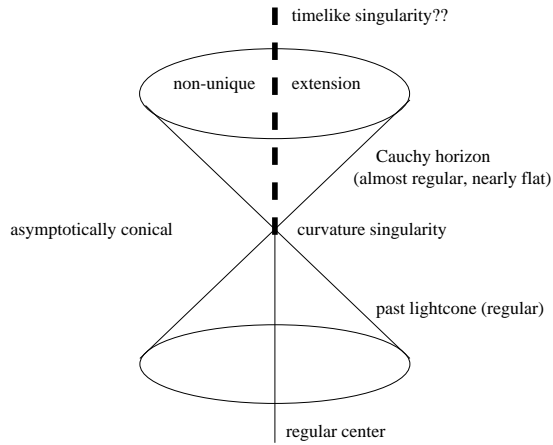}}
\caption{The global structure of spherically symmetric critical spacetimes. One
dimension in spherical symmetry has been suppressed.} 
\end{figure}




\end{document}